\documentclass[11 pt]{article}
\pdfoutput=1 
\usepackage{jheppub} 
\usepackage{orcidlink}
\usepackage{lscape}
\usepackage{multicol}
\usepackage{mathrsfs}  
\usepackage{xcolor}
\usepackage{color, colortbl}
\usepackage{empheq}
\usepackage{tabularx}
\usepackage{adjustbox}
\usepackage{physics}
\usepackage{caption}
\usepackage{subcaption,longtable,stmaryrd}
\usepackage{slashed}
\usepackage{bigints}
\usepackage{cancel}
\usepackage{multicol}
\usepackage{blindtext}
\usepackage{tikz}
\usepackage{enumitem}
\usepackage[customcolors,shade]{hf-tikz}
\usepackage{graphicx}
\usepackage{changepage}
\usepackage{cleveref}
\usepackage[most]{tcolorbox}
\usepackage{cancel}
\DeclareSymbolFont{usualmathcal}{OMS}{cmsy}{m}{n}
\DeclareMathAlphabet\mathbfcal{OMS}{cmsy}{b}{n}
\DeclareSymbolFontAlphabet{\mathcal}{usualmathcal}
\DeclareSymbolFont{rmlargesymbols}{OMX}{mdbch}{m}{n}
\DeclareMathSymbol{\rmintop}{\mathop}{rmlargesymbols}{82}
\DeclareMathSymbol{\rmointop}{\mathop}{rmlargesymbols}{72}

\definecolor{mygray}{gray}{0.5}

\title{\boldmath {{ Finite cutoff JT gravity: Baby universes, Matrix dual, and (Krylov) Complexity}}}

\author[a]{Arpan Bhattacharyya,}
\author[a]{Saptaswa Ghosh,}
\author[a]{Sounak Pal}
\author[a,b]{and Anandu Vinod}
\affiliation[a]{\it Indian Institute of Technology, Gandhinagar, Gujarat-382055, India}
\affiliation[b]{\it Indian Institute of Science Education and Research Kolkata, West Bengal-741246, India}
\emailAdd{abhattacharyya@iitgn.ac.in}
\emailAdd{saptaswaghosh@iitgn.ac.in}
\emailAdd{palsounak@iitgn.ac.in}
\emailAdd{av20ms164@iiserkol.ac.in}

\abstract{In this paper, as an application of the `Complexity = Volume' proposal, we calculate the growth of the interior of a black hole at late times for finite cutoff JT gravity. Due to this integrable, irrelevant deformation, the spectral properties are modified non-trivially. The Einstein-Rosen Bridge (ERB) length saturates faster than pure JT gravity. We comment on the possible connection between Krylov Complexity and ERB length for the deformed theory. Apart from this, we compute the emission probability of baby universes in the deformed theory and \textcolor{black}{find that it changes due to the deformation parameter only if we turn on Lorentzian evolution}. We also find that the saturation time of the deformed theory relative to the undeformed one depends on the inverse temperature. We also highlight the subtleties involved in the dual matrix model and comment on the possible one-cut universality. Finally, we comment on the possible correction to the volume of the moduli space arising from the non-perturbative correction of the spectral curve induced by the finite boundary cutoff.}  
\begin{document}
\maketitle
\section{Introduction}

Although a microscopic understanding of the universe has eluded fundamental physics for decades, accelerated progress has been made in recent times in the context of the AdS/CFT correspondence \cite{Maldacena:1997re,Witten:1998qj}.  There is a renewed interest in lower-dimensional models of quantum gravity in two dimensions \cite{Jackiw:1984je, Teitelboim:1983ux,Almheiri:2014cka} as well as in three dimensions, pioneered in \cite{Maloney:2007ud,Yin:2007gv} and more recently \cite{Collier:2023fwi,deBoer:2024mqg, Collier:2024mgv,Bhattacharyya:2024vnw,Post:2024itb,Takahashi:2024ukk}.  Particularly in two-dimensions, Jackiw-Teitelboim (JT) gravity \cite{Jackiw:1984je,Teitelboim:1983ux}\footnote{Refer \cite{Mertens:2022irh,Moitra:2019bub,Moitra:2022glw} for detailed reviews.}, a model of dilaton gravity in a 2D Euclidean spacetime of negative cosmological constant.  Its connection to the Sachdev-Ye-Kitaev (SYK) model \cite{Sachdev:1992fk,2015escq.progE...2K, 2015escq.progE..38K,Sachdev:2015efa} has helped characterize its chaotic nature and has become a topic of active research interest \cite{Shenker:2014cwa,Maldacena:2015waa,Stanford:2015owe,Cotler:2016fpe,Jensen:2016pah,Engelsoy:2016xyb,Chowdhury:2017jzb,Altland:2022xqx}.  This connection is enabled by the fact that the low-energy dynamics of the SYK model are described by the 1D Schwarzian theory, which in turn is the boundary description of bulk 2D JT gravity \cite{2015escq.progE...2K, 2015escq.progE..38K,Maldacena:2016hyu,Maldacena:2016upp,Kourkoulou:2017zaj,Kitaev:2017awl,Mertens:2018fds,Lin:2019qwu}. 
This has also led to the study of gravity in the context of random matrix theories \cite{Cotler:2016fpe,Saad:2018bqo}.  The seminal work of Saad-Shenker-Stanford \cite{Saad:2019lba}, has further established the connection between JT gravity, matrix integrals, and topological recursion \cite{Stanford:2019vob,Eynard:2007fi}.\\

\noindent
The study of perturbations of black holes has taken a central role in key problems such as the black hole information paradox \cite{Hawking:1976ra,Mathur:2009hf,Raju:2020smc} since the introduction of the AdS/CFT duality \cite{Maldacena:2001kr}.  Two-point functions of quantum fields outside the eternal black hole, widely separated in time, are used to study perturbations in the black hole.  \textcolor{black}{While semiclassical analysis suggests a two-point function that decays forever, such functions for the boundary theory on a compact space with a discrete spectrum point to saturation at late times} \cite{Maldacena:2001kr,Goheer:2002vf,Dyson:2002pf,Barbon:2003aq}.  The precise analytic description of the two-point function is challenging to obtain since the fluctuations around the late-time average are highly erratic and sensitive to the fine details of the energy spectrum \cite{Barbon:2003aq,prange1997spectral}.  This behaviour can be studied more readily by recasting it as a problem in random matrix theory and the SYK model.  Such studies indicate that the two-point function decays initially, then shows a period of linear growth, called the `ramp' until the growth finally stops at a time exponentially long in the entropy, which is referred to as the `plateau' \cite{Cotler:2016fpe}.\\

\noindent
This non-decaying behaviour of correlation functions in JT gravity \cite{Maldacena:2016upp,Yang:2018gdb,Gross:2017aos,Lam:2018pvp,Mertens:2017mtv,Blommaert:2018oro,Blommaert:2019hjr,Bulycheva:2019naf,Iliesiu:2019xuh} is attributed to the topology change due to Euclidean wormholes on the bulk \cite{Lavrelashvili:1987jg,HAWKING1987337,GIDDINGS1988890,Coleman:1988cy,COLEMAN1988643,GIDDINGS1989481,KLEBANOV1989665,Maldacena:2004rf,Arkani-Hamed:2007cpn}.  This is regarded as a tunneling process in which an asymptotically AdS `parent universe' emits or absorbs a closed `baby universe'.  Such baby universes can also form `loops' when they are reabsorbed after getting emitted, or they can end in a `D-brane' state.  The behaviour of the spectral form factor \cite{Okuyama:2023pio,Anegawa:2023klh}, can also be attributed to this process \cite{Saad:2018bqo,Saad:2019lba}.  Topology-changing effects play an important role in correlation functions since a very large parent universe can become small by emitting a very large baby universe and can be reabsorbed by the parent universe with a non-decaying probability.  Emission and absorption probabilities of baby universes in JT gravity and its relations to the Eigenstate Thermalization Hypothesis (ETH) \cite{Srednicki:1994mfb,Deutsch:1991msp} were presented in \cite{Saad:2019pqd}, which are of particular interest for us.\\

\noindent
Given the fact that the computations were done mostly for pure JT gravity, it is tempting to study integrable irrelevant deformations, which change the UV behaviour of the theory in a non-trivial way, and see whether the above observations still hold.  One such example is $T\bar{T}$ deformation \cite{Zamolodchikov:2004ce,Smirnov:2016lqw,Cavaglia:2016oda} in the family of integrable deformations.  \textcolor{black}{Irrelevant deformations provide a straightforward handle to examine how microscopic
couplings influence nonperturbative features.  Recent studies of SYK and related
models indicate that wormhole and
half-wormhole saddles, as well as interior-reconstruction diagnostics, persist under
such deformations, and that factorization depends on boundary couplings
\cite{Mukhametzhanov:2021nea,Das:2022uhj,Almheiri:2021jwq}.  In two dimensions, the
$T\bar T$ deformation is a standard and tractable example.  It allows us to track
how partition functions, spectral data, and semiclassical saddles change as the
irrelevant coupling varies.  This work uses this setting to relate the SYK lessons to
a continuum QFT and to clarify when factorization, coupling sensitivity, and
ensemble-type interpretations appear beyond discrete models}.  Apart from that, there are several studies of  $T\bar{T}$ deformation have been done in the context of field theory and gravity \cite{Conti:2018jho,Conti:2018tca,Conti:2019dxg,Jafari:2019qns,Guica:2022gts,Morone:2024ffm, Bielli:2024khq,Chang:2024voo,Tsolakidis:2024wut,He:2024pbp,Babaei-Aghbolagh:2025lko} \footnote{The list is by no means exhaustive, interested readers are referred to the review \cite{Jiang:2019epa} and the references there in.} .  $T\bar{T}$ is a composite operator made up of the holomorphic and anti-holomorphic parts of the stress tensor.  One uses the point-splitting method to make sense of such operators in $d>1$.  However, in $d = 1$, in the absence of a spatial direction, they are still well-defined.  By applying a $T\bar{T}$ deformation to the boundary Schwarzian theory (in 1D) of JT gravity, we study the deformed spectrum and compute various physical quantities like the growth of wormhole length in the dual bulk theory, and study its saturation properties.  These have a deep connection with the notion of chaos and complexity.  As a test of the `Complexity $=$ Volume' conjecture \cite{Stanford:2014jda,Susskind:2014rva}, we try to compute and see the nature of the saturation of complexity, which is basically equal to the growth of the interior, in such deformed theories.  $T\bar{T}$ deformation changes the energy spectrum non-trivially.  This deformation mimics the insertion of a brane in some sense.  Like in the presence of a End-of-World (EOW) brane \cite{Alishahiha:2022kzc,Blommaert:2020hgi,Gao:2021uro} some of the properties like the probabilities of emission of a baby universe and the rate of saturation of complexity changes in a similar way for $T \bar{T}$ deformation.  \textcolor{black}{The dual matrix model potential shows natural minima and slow oscillation at a specific value depending on the cut-off parameter $\lambda$} \cite{Rosso:2020wir}.  For JT gravity, the matrix potential shows erratic oscillation, and one needs to insert the notion of { ZZ instantons} \cite{Saad:2019lba}, which gets regulated quite naturally for the deformed ($T\bar{T}$) theory as we will see in this paper.  This makes this study all the more interesting.  Recently, different aspects like free energy, SFF, etc., have been calculated in such a setup \cite{Ebert:2022gyn, Bhattacharyya:2023gvg}.  Apart from this, $T\bar{T}$ deformed entanglement entropy and complexity has been explored in \cite{Donnelly:2018bef,Banerjee:2019ewu,He:2019vzf,He:2022ryk,FarajiAstaneh:2024fpv,Chattopadhyay:2024pdj}. \\

\noindent
\textcolor{black}In this paper, beside computing the emission probability of baby universe in presence of $T\bar{T}$ deformation,  we also {\textit{demonstrate that the saturation time of wormhole growth in the deformed theory, relative to the undeformed case, is not universal but depends on the value of the inverse temperature, \(\beta\).  In particular, we find that up to a certain value of \(\beta\), the growth of the deformed length saturates faster than in the undeformed theory.  Beyond this point, a transition occurs, after which the undeformed theory saturates more rapidly.}} We also comment on the possible connection with {Krylov complexity}, which is a notion of complexity related to operator growth \cite{PhysRevX.9.041017,Barbon:2019wsy,Rabinovici:2020ryf}. The connection between Krylov complexity in the ensemble dual of JT gravity to volume growth has been shown in \cite{Kar:2021nbm,Jian:2020qpp,Rabinovici:2023yex,Xu:2024gfm,Balasubramanian:2024lqk,Ambrosini:2024sre} \footnote{Authors of \cite{Rabinovici:2023yex,Ambrosini:2024sre}, related the Krylov complexity with the length of the Einstein-Rosen Bridge using the construction of \cite{Lin:2022rbf,Xu:2024gfm} for JT gravity which is subsequently extended for certain other 2D gravity model \cite{Heller:2024ldz}.  For more details, interested readers are referred to  \cite{Kar:2021nbm,Rabinovici:2023yex,Xu:2024gfm,Ambrosini:2024sre, Heller:2024ldz}. }. We show that this connection is valid even after the introduction of $T\bar{T}$ deformation.  To show this, one first computes the two-point function of primary insertions and then can read off the \textit{Lanczos coefficients $(b_n)$} by calculating the $n$-th order moments and constructing a determinant of the moment matrix \cite{PhysRevX.9.041017,viswanath1994recursion,Avdoshkin:2019trj}.  The growth of the $b_n$ is an inherent behaviour of quantum chaos \cite{PhysRevX.9.041017}.  The volume of the maximal slices in Euclidean geometries initially grows exponentially faster, then linearly, mimicking the Krylov complexity.  But if we only consider the wormhole contribution, the volume of the maximal slice does not saturate (which will require a sum of all possible geometries). 
 We discuss some of the aspects of Krylov complexity and black-hole interior growth for our deformed theory.\\ 

\noindent
 \textcolor{black}{The paper is organized as follows:  In section (\ref{sec2}), we briefly review the boundary particle formalism in JT-gravity, which is relevant for the computation of the Hartle-Hawking state in $T\bar{T}$ deformed theory.  Then, in sections (\ref{sec3}) and (\ref{sec4}), we discuss how one can arrive at the expression for the $T\bar{T}$ deformed propagator and discuss the computation of the emission probability of baby universes for the deformed theory.  In section (\ref{sec5new}), we compute the double trumpet function, which is required for the computation of the planar two-loop resolvent needed for the density-density correlator.  In section (\ref{sec5}), we briefly discuss the nature of the possible $T\Bar{T}$  deformed matrix model and comment on its cut structure.  We also compute the density correlators in this section.  Further, in section (\ref{sec6}), we compute how the growth of ERB changes with time compared to its behavior with that of undeformed JT gravity.  As an application of the \textit{`Complexity = Volume'} conjecture, we compute the expectation of length in the deformed theory and find how it deviates from pure JT gravity and its dependence on the inverse temperature.  Again this is one of the main result of the pape. Finally, in section (\ref{sec7}), we summarize our main findings and conclude with some future directions.  We briefly review the partition function and density of state computation in Appendix~\eqref{aa1}.  Some details regarding the computations of the deformed moduli space volume using the spectral curve have been given in Appendix~(\ref{a1}).  We also comment on why we expect the $T\bar{T}$ deformed moduli space volume to be changed and present some of the relevant discussions in this specific context.}

\section{Boundary particle formalism for JT gravity }\label{sec2}

For Euclidean JT gravity minimally coupled to the matter sector, the action is given by,

\begin{align}
\begin{split}&
\mathcal{S}[g,\Phi,\phi]=-S_0\chi+S_{\text{JT}}[g,\Phi]+S_{\text{matter}}[g,\phi]\,,\\&
S_{\text{JT}}=-\int_{\mathcal{M}}\sqrt{g}\Phi(R+2)-2\int_{\partial \mathcal{M}}\sqrt{h}\Phi(K-1)\,.
\end{split}
\end{align}
We have set $16\pi G_{N}=1$ and $S_{\text{matter}}$ is the action of matter QFT. $\chi$ denotes the Euler-characteristic of the manifold and $S_0$ is the entropy.  At the AdS boundary, 
\begin{align}
    g_{\tau\tau}|_{\partial \mathcal{M}}=\frac{1}{\epsilon^2}\,,\quad   \Phi|_{\partial \mathcal{M}}=\frac{\phi_b}{\epsilon}\,.
\end{align}
We work in the $\epsilon\rightarrow 0$ limit. $\frac{1}{\phi_b}$ plays the role of $G_N\,$ The semiclassical limit $\phi_b\rightarrow \infty$ corresponds to the large-N limit of dual CFTs in the framework of holography. Now, the boundary particle formalism is defined by the path integral \cite{Yang:2018gdb},
\begin{align}
    K_{\beta}(\phi_2,\psi_2,\phi_1,\psi_1)=\int \mathcal{D}\phi \mathcal{D}\psi \mathcal{D}\pi_{\phi} \mathcal{D}\pi_{\psi} \,\textrm{exp}\,\Bigg(\int_0^\beta d\tau \Bigg[i\pi_{\psi}\psi'+i\pi_{\phi}\phi'-\frac{1}{2\phi_b}\Bigg[\frac{\pi_{\psi}^2}{2}+i \pi_{\phi}e^{\psi}-\frac{1}{2}e^{2\psi}\Bigg]\Bigg)\label{2.4y}
\end{align}
where \textcolor{black}{$\phi,\psi$ are angular and radial coordinates respectively}. This can be viewed as a worldline theory where the AdS boundary plays the role of the target space. $\phi$ is chosen to be a non-compact bosonic field during the path integral. The canonical Hamiltonian is given by
\begin{align}
    2\phi_b H=\frac{\pi_{\psi}^2}{2}+i\pi_{\phi} e^{\psi}-\frac{1}{2}e^{2\psi}
\end{align}
and (\ref{2.4y}) can be written as,
\begin{align}
\begin{split}
    K_{\beta}(\phi_2,\psi_2,\phi_1,\psi_1):=\langle\phi_2\psi_2|e^{-\beta H}|\phi_1\psi_1\rangle,
    \end{split}
    \end{align}
    with the property,
    \begin{align}
    \langle \phi_2\psi_2|\phi_1\psi_1\rangle=\delta(\phi_2-\phi_1)\delta(\psi_2-\psi_1).
    \end{align}
    Furthermore, one should note that the field $\pi_{\phi}$ acts as the Lagrange multiplier which imposes the constraint 
    \begin{align}
        \phi'=\frac{e^{\psi}}{2\phi_b}>0\,.    \end{align}
This implies that the particle's trajectory doesn't intersect itself as $\phi$ increases monotonically with time. Given this formalism, 
    in the following section, we discuss $T\bar{T}$ deformations and how to calculate the deformed propagator. 
    
\section{$T\Bar{T}$ deformation and JT gravity}\label{sec3} 

$T\bar{T}$ deformation is an integrable irrelevant deformation. 
The composite operator $T\bar{T}$ in two-dimensional quantum field theory is constructed from the chiral components $T$ and $\bar{T}$ of the energy-momentum tensor $T_{\mu\nu}$. It changes the ultraviolet behaviour of the theory in a crucial way.
Denoting the complex coordinates as $z,\bar z$, we can write the chiral components of the energy-momentum tensor as follows,
\begin{equation}
   T= -2\pi T_{zz},\,\,\,\,\,\,\,\,\, \bar T= -2\pi T_{\bar z\bar z},\,\,\,\,\,\,\,\,\, \Theta =2 \pi T_{z\bar z}\,.
\end{equation}
This yields the following \cite{Cavaglia:2016oda, Smirnov:2016lqw},
\begin{equation}\langle T \bar T\rangle= \langle T\rangle \langle  \bar T\rangle-\langle\Theta\rangle^2\,.\end{equation}
In the limit $z\rightarrow z'$, \,  $T(z)\bar{T}(z')$ and $\Theta(z)\bar{\Theta}(z')$ are individually divergent. But the combination $T(z)\bar{T}(z') - \Theta(z){\Theta}(z')$  is finite as there is a fine cancellation of the divergence. 
Now the flow equation is given by \cite{Cavaglia:2016oda, Smirnov:2016lqw, Zamolodchikov:2004ce},
\begin{align}
    \frac{\partial S_{E}}{\partial \lambda}=8\int d^2x \sqrt{\gamma}\, T\bar{T}\,.
\end{align}
We have,
$T_{\mu}^{\mu}=-16\lambda T\bar{T}=-2\lambda (T_{ij}T^{ij}-(T_{i}^{i})^2)$ and one can solve it to get,

\begin{align}
    T_{\phi}^{\phi}=\frac{T^{\tau}_{\tau}+4\lambda T_{\tau\phi}T^{\tau\phi}}{4\lambda T_{\tau}^{\tau}-1}\,.
\end{align}
\noindent
Now, using $\langle T_{\tau}^{\tau}\rangle =E$ and $\langle T_{\tau\phi}\rangle=\langle T^{\tau\phi}\rangle=iJ$, one achieves the following differential equation for the energy levels,
\begin{align}
    \frac{\partial E}{\partial \lambda}=\frac{E^2-J^2}{1/2-2\lambda E}
\end{align}
and its solution can be written as \cite{Zamolodchikov:2004ce},
\begin{align}
 E\to   {E}(\lambda)=\frac{1}{4\lambda}\left(1-\sqrt{1-8\lambda E+16\lambda^2 J^2}\right)\,. \label{deformed}
\end{align}
In the context of holography, the deformed spectrum in (\ref{deformed}) agrees with that of finite cut-off two-dimensional black holes upon the identification $|\lambda|=2\pi G_N/r_c^2$ \cite{Iliesiu:2020zld}. The next section is devoted to the computation of $T\bar{T}$ deformed propagator in JT gravity, and we use that propagator to compute the deformed Hartle-Hawking state required for subsequent computations.

\subsection{The $T\bar{T}$ deformed propagator}

Before proceeding further, in this section, we review the derivation of the $T \bar{T}$ deformed propagator in JT gravity, which will be useful for the computation of the Hartle-Hawking wave function. The expression for the undeformed propagator is given in \cite{Chakraborty:2020xwo}. To calculate the propagator for the deformed theory, we start with deforming the boundary theory using the following kernel \cite{Chakraborty:2020xwo}.
\begin{align}
 K(\beta,\beta') ={\frac{\beta}{\sqrt{-8\pi\lambda\beta'^{3}}}e^{\frac{(\beta-\beta')^2}{8\lambda \beta'}}}\,.  \label{deformed1}
\end{align}
Here, $\lambda$ is the constant parameter and denotes the coupling of $T\bar{T}$ deformation. There are two signs for $\lambda$ \cite{Ebert:2022gyn}\,. $\lambda<0$ is called the \textit{good sign} and for  $\lambda>0$ the unitarity is violated when $\lambda>\frac{1}{8E}$. It is sometimes called the  \textit{bad sign}. The expression mentioned in (\ref{deformed1}) is for the good sign.   One can also proceed to compute the $T\bar{T}$ deformed propagator from bulk calculations by adding a finite cut-off boundary term, as the holographic dual to the $T\bar{T}$ deformation corresponds to adding a finite cut-off surface to the dual geometry \cite{McGough:2016lol}. The variational principle can be made well-defined by considering the Dirichlet boundary condition \cite{Iliesiu:2020qvm}.\\

\noindent Now, we proceed to calculate the $T\bar{T}$ deformed propagator. \textcolor{black}{Using (\ref{2.4y}), we can find the propagator for the deformed Schwarzian theory. This is given by,
\begin{align}
\begin{split}\mathcal{K}_{\beta}(\phi_2,\psi_2,\phi_1,\psi_1)&=\exp\Big( (e^{\psi_2}+e^{\psi_1})\cot\Big(\frac{(\phi_1-\phi_2)}{2}\Big)\Big)\\&\times \frac{1}{\pi^2 \sin\frac{(\phi_2-\phi_1)}{2}}\int_0^\infty ds\, s\, \sinh(2\pi s)\,e^{\frac{-\beta s^2}{4\phi_b}}\,K_{2is}\Bigg(\frac{2e^{\frac{\psi_1+\psi_2}{2}}}{\sin\frac{(\phi_2-\phi_1)}{2}}\Bigg)
\label{eq:undef_prop}
\end{split}
\end{align}
where $\phi$ and $\psi$ are angular and radial coordinate respectively. $\beta$ is the length of the thermal circle. The propagator denotes the propagation of the boundary edge mode between the coordinates $(\phi_1,\psi_1)$ to $(\phi_2,\psi_2)$.} Now for the $T\bar{T}$ deformed theory, the propagator after the kernel integration using (\ref{deformed1}) becomes \cite{Ebert:2022gyn},

\hfsetfillcolor{gray!8}
\hfsetbordercolor{white}
\begin{align}\begin{split}
\tikzmarkin[disable rounded corners=true]{r}(0.25,-1)(-0.3,1.0) \mathcal{K}_{\beta,\lambda}&=\int_0^\infty d\beta'K(\beta,\beta')\mathcal{K}_{\beta'}(\phi_2,\psi_2,\phi_1,\psi_1)\\&
=\exp\Big( (e^{\psi_2}+e^{\psi_1})\cot\Big(\frac{(\phi_1-\phi_2)}{2}\Big)\Big)\frac{1}{\pi^2 \sin\frac{(\phi_2-\phi_1)}{2}}\\&\times \int_0^\infty\, ds \,s \sinh(2\pi s)K_{2is}\left(\frac{2e^{\frac{\psi_1+\psi_2}{2}}}{\sin\frac{(\phi_2-\phi_1)}{2}}\right)\text{exp}\left[{\frac{\beta}{4 \lambda }  \left(\sqrt{1-\frac{2 \lambda  s^2}{\phi_b }}-1\right)}\right]\label{3.9o}\,.
\tikzmarkend{r}\\ 
 \end{split} 
\end{align}\\

\noindent \textcolor{black}{The expression is valid for $\lambda<0$} and is equivalent to the undeformed propagator for $\lambda \rightarrow 0$.  As the flow equation is the same for both the bulk and boundary \cite{Chakraborty:2020xwo}, we will use the same kernel, as shown in (\ref{3.9o}), used to deform the boundary theory to deform the bulk partition function. 

To compute the  Hartle-Hawking state, which is dual to a thermofield double state for double-sided black holes, we need one more ingredient, namely, the density of states. We will discuss that in the next sub-section. 

\subsection{Bulk partition function dual to $T\bar{T}$ deformed theory and density of states}

In two dimensions, all the topologies that contribute to the gravitational path integral are built by gluing some basic building blocks like disks and trumpets. To construct the gravitational path integral, we need to integrate over the bulk moduli space and the boundary wiggles. Boundary wiggles are given by diffeomorphisms (Diff($S^1$)), and we need to integrate over this large diffeomorphism, which preserves the boundary. Apart from this, one also needs to integrate over the moduli space of underlying Riemann surfaces. In principle, one needs to sum over all topologies to recover the non-perturbative result for the one-point or connected two-point functions of the partition function, i.e. $\langle Z(\beta)\rangle$ or $\langle Z(\beta_1)Z(\beta_2)\rangle$. However, in general, the sum is very tough to perform because of the structure of the complicated hyperbolic moduli space volumes, which obey {\textit{Mirzakhani recursion relations}}. For further details, readers are referred to Appendices~(\ref{aa1}) and (\ref{a1}) for further details regarding computation of the  $\langle Z(\beta)\rangle$ for $T\bar{T}$ deformed case and comments on possible deformation of the volume of moduli space for negative sign of the deformation parameter.  \\

\noindent
 The $g$-genus and $n$-boundary  partition function for the undeformed JT gravity can be obtained by integrating over the dilaton field and the metric,
  \begin{align}
      \begin{split}
          Z_{g,n}(\beta_1,..,\beta_n)=\int \frac{\mathcal{D}\varphi \, \mathcal{D}g_{\mu\nu}}{\text{Vol(diff.)}} \,e^{-S_{\text{JT}}[g,\varphi]}\,.
      \end{split}
  \end{align}
  \noindent
Here, $\frac{\mathcal{D}g_{\mu\nu}}{\text{Vol.(diff.)}}$ is the diffeomorphism invariant measure.
The disk and the trumpet partition functions for JT gravity can be computed using a Schwarzian theory, and they are given below as follows \cite{Stanford:2017thb,Saad:2019lba},
\begin{align}
    \begin{split}
    \label{jt_part}
        Z_0^{\text{disk}}(\beta) &= \begin{minipage}[h]{0.15\linewidth}
	\vspace{4pt}
	\scalebox{0.7}{\includegraphics[width=\linewidth]{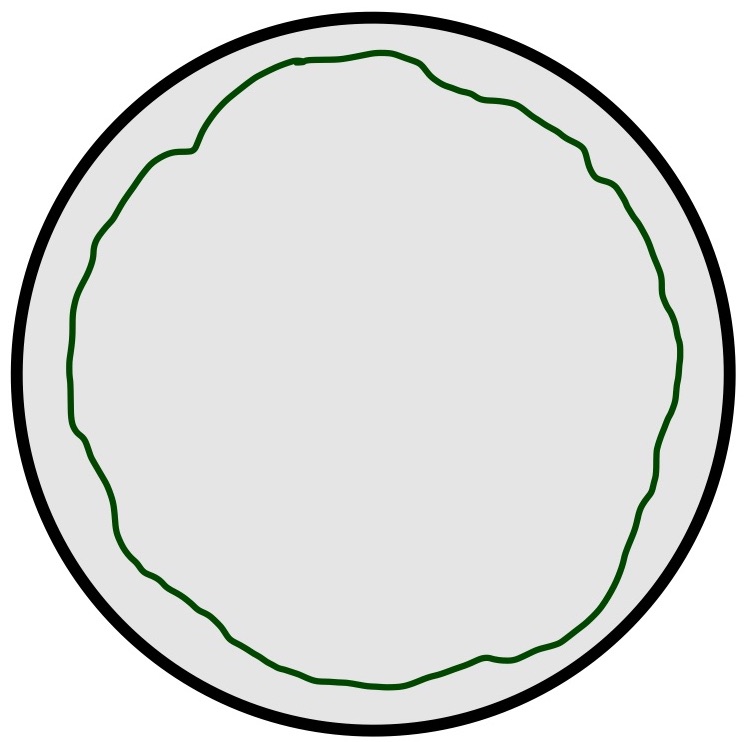}}
\end{minipage}\hspace{-0.59cm}=\frac{e^{\frac{\pi^2}{\beta}}}{4\sqrt{\pi}\beta^{\frac{3}{2}}},\hspace{1.5cm}Z_0^{\text{trumpet}}(b,\beta) =\begin{minipage}[h]{0.15\linewidth}
	\vspace{4pt}
	\scalebox{0.9}{\includegraphics[width=\linewidth]{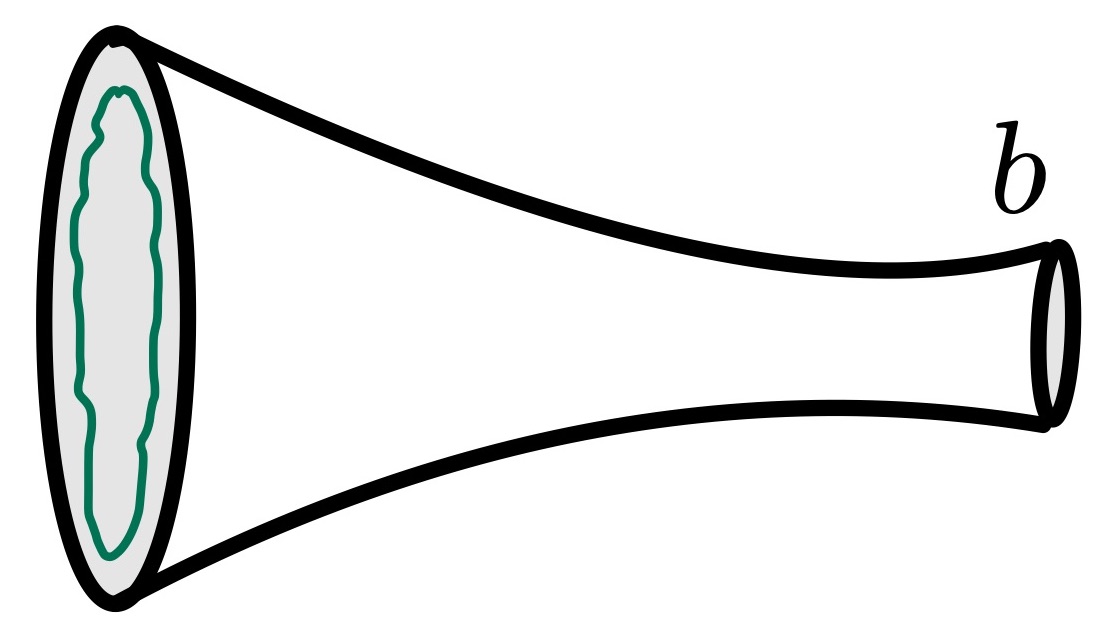}}
\end{minipage} =\frac{e^{-\frac{b^2}{4\beta}}}{2\sqrt{\pi\beta}}\,.
    \end{split}
\end{align}
Here, $b$ denotes the length of the geodesic boundary. Furthermore, an inverse Laplace transform of these partition functions results in the expressions for the corresponding undeformed density of states as a function of energy $E\,.$
\begin{align}
    \begin{split}
        \rho_0^{\text{disk}}(E) &= \frac{\sinh(2\pi\sqrt{E})}{4\pi^2},\hspace{1cm}\rho_0^{\text{trumpet}}(b,E) = \frac{\cos(b\sqrt{E})}{2\pi\sqrt{E}}\,.\label{3.2 m}
    \end{split}
\end{align}
We know that the deformed spectrum is given as \cite{Gross:2019ach}:
\begin{equation}
    \mathcal{E}(\lambda) = \frac{1}{4\lambda}(1-\sqrt{1-8\lambda E})\label{3.15k}\,.
\end{equation}
Note that the limit $\lambda \rightarrow 0$ leads to the undeformed spectrum. The deformed density of states can then be written as,
\begin{align}
\begin{split}
    \rho_\lambda(\mathcal{E})& = \frac{dE(\mathcal{E})}{d\mathcal{E}}\,\rho_0 (E(\mathcal{E})),
    \end{split}
\end{align}
where $\rho_0$ is the undeformed density of states and
\begin{align}\begin{split}\label{4.4l}
    E(\mathcal{E}) = \mathcal{E}(1-2\lambda\mathcal{E}).
\end{split}
\end{align}
Hence, we can write the deformed density of states for the disk and trumpet geometries as
\begin{align}
\begin{split}
    \rho_\lambda^{\text{disk}}(E) = (1-4\lambda E)\frac{\sinh\left(2\pi\sqrt{E(1-2\lambda E)}\right)}{4\pi^2}\,,\label{3.16y}
    \end{split}
\end{align}
\begin{align}
    \begin{split}
    \rho_\lambda^{\text{trumpet}}(b,E) = (1-4\lambda E)\frac{\cos\left(b\sqrt{E(1-2\lambda E)}\right)}{2\pi\sqrt{E(1-2\lambda E)}}\,.
    \end{split}
\end{align}
\textcolor{black}{We now have all the necessary ingredients to compute the Hartle-Hawking wavefunction and, consequently, the rate of baby-universe emission during the Lorentzian evolution of the wormhole length. This is one of the main computations done in this paper.}
\section{Hartle-Hawking wave function and baby universes }\label{sec4}  

\begin{figure}[t]
    \centering
\scalebox{0.7}{\includegraphics[width=0.5\linewidth]{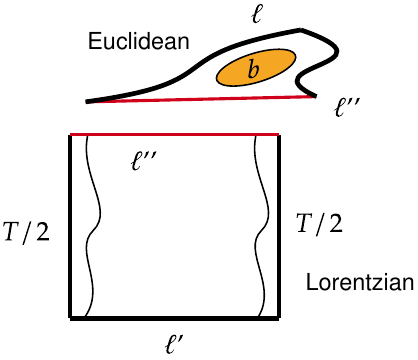}}
    \caption{Figure describing the Euclidean and Lorentzian parts of emission of the baby universe.}
    \label{fig:2}
\end{figure}

In this section, we compute the Hartle-Hawking wave function, which is the primary ingredient to compute the emission probability of baby universes (refer to Fig.~(\ref{fig:2}) for a representative diagram). \textcolor{black}{We will eventually show that the emission amplitude of baby universes is greater in $T\bar T$ deformed theory for the \textit{good sign} of deformation parameter.} The deformed Hartle-Hawking wavefunction in the length basis, considering contributions only from the disk geometry, can be computed using the deformed propagator \eqref{3.9o} as done for JT gravity in \cite{Kolchmeyer:2023gwa}. We start with the following:
\begin{align}
    \begin{split}
\mathcal{K}_{\beta,12,\lambda} &= e^{\frac{\psi_1 + \psi_2}{2}}\mathcal{K}_{\beta,\lambda}(\phi_2,\psi_2,\phi_1,\psi_1)\,.
    \end{split}
\end{align}
For the half disk geometry, we set $\phi_1 =0,\phi_2 = \pi$ and perform the following coordinate transformations
\begin{align}
    \begin{split}
        e^{-\frac{\ell}{2}} = \frac{e^{\frac{\psi_1 + \psi_2}{2}}}{2\sin(\frac{\phi_2-\phi_1}{2})},\hspace{1cm}s^2=E'
    \end{split}
\end{align}
where $l$ is the renormalized length on the hyperbolic disk between points $(\phi_1,\psi_1)$ and $(\phi_2,\psi_2)$. \textcolor{black}{The propagator uses a convention where $\beta$ is the AdS boundary length, we need to rescale $\beta \rightarrow \beta/2$ for our case and also set $\phi_b = 1/2$. Both of these can be done by the single rescaling $\lambda\rightarrow 2\lambda$}. We finally arrive at an expression for the Hartle-Hawking wavefunction in the length basis for the deformed theory.

\begin{equation} \psi^{\text{disk}}_{\lambda,\beta/2}(\ell)\equiv \langle{\ell}|{HH_{\beta/2,\lambda}} \rangle=\frac{1}{\pi^2}\int_0^\infty  dE' \sinh(2\pi\sqrt{E'})e^{-\frac{\ell}{2}}K_{2i\sqrt{E'}}(4e^{-\frac{\ell}{2}})e^{\frac{\beta}{8\lambda}(\sqrt{1-8\lambda E'}-1)}\,.
\end{equation}\\
Performing a variable change $E' =  E(1-2\lambda E)$ we have,
\begin{equation}
    \psi^{\text{disk}}_{\lambda,\beta/2}(\ell) = \int_0^\infty dE \rho^{\text{disk}}_\lambda(E)e^{-\frac{\beta E}{2}}\left(4e^{-\frac{l}{2}}K_{2i\sqrt{E(1-2\lambda E)}}(4e^{-\frac{l}{2}})\right)\,.
\end{equation}
Defining
\begin{equation}
    \psi_{E,\lambda}(\ell) := \langle{\ell}|{E} \rangle\equiv 4e^{-\frac{\ell}{2}}K_{2i\sqrt{E(1-2\lambda E)}}(4e^{-\frac{\ell}{2}})\,,
\end{equation}
where $|E\rangle$ denotes the bulk eigenstates. The overlap of two Hartle-Hawking wavefunctions for the deformed theory is defined as
\begin{equation}
    \langle{HH_{\beta/2}}|{HH_{\beta'/2}}\rangle \equiv \int_{-\infty}^\infty e^{\ell} d\ell \psi^{*\,\,\text{disk}}_{\lambda,\beta/2}(\ell)\psi^{\text{disk}}_{\lambda,\beta'/2}(\ell)\,.
\end{equation}
The normalization of the wavefunction for pure JT gravity in the energy basis is given by,
\begin{align}
   \int_{-\infty}^{\infty}e^\ell d\ell \psi_E^{*}(\ell)\,\psi_{E'}(\ell)=\frac{\delta(E-E')}{\rho^{\text{disk}}_0(E)}
\end{align}
where $\rho^{\text{disk}}_0(E)$ is defined in \eqref{3.2 m}.
Similarly, for $T\bar{T}$ case, deformed wavefunctions satisfy the normalization \footnote{This can be seen by simply rescaling. $E\rightarrow E(1-2\lambda E)\,.$}
\begin{align}
    \int_{-\infty}^\infty e^\ell d\ell\, \psi^{*}_{E,\lambda}(\ell)\psi_{E',\lambda}(\ell)=\frac{\delta(E-E')}{\rho^{\text{disk}}_\lambda(E)},
\end{align}
This expression is essential to evaluate the norm of the Hartle-Hawking wavefunctions.

\subsection*{Trumpet wavefunction and Baby universes:}
\begin{figure}[htb!]
\centering
\scalebox{0.18}{\includegraphics{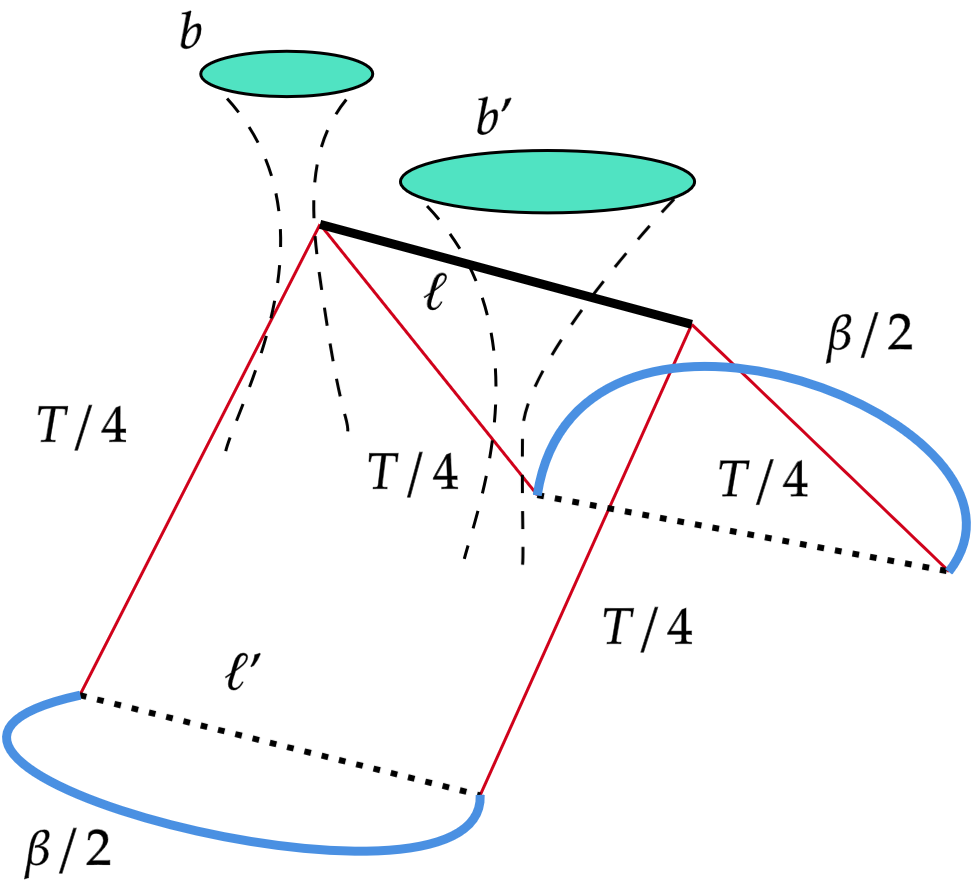}}
\caption{The growth of ERB length and the emission of baby universes are shown in the figure.}\label{fig12}
\end{figure}

\begin{figure}
    \centering
\includegraphics[width=0.5\linewidth]{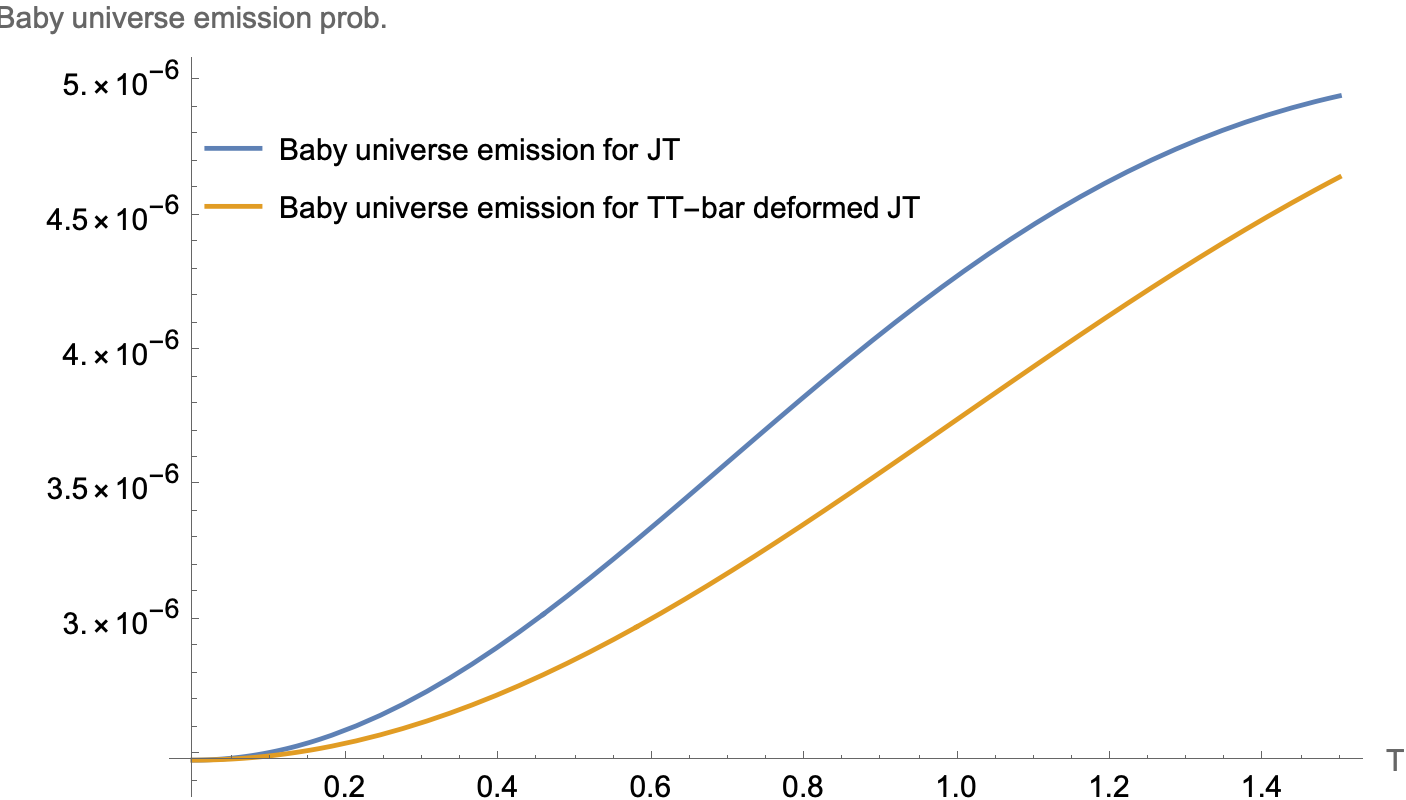}
    \caption{Schematic diagram showing the probability of emission of baby universes for JT gravity and $T\bar{T}$ deformed theory. The yellow graph is plotted for $\lambda=-0.1$. }
    \label{fig:2a}
\end{figure}

The trumpet wavefunction for pure JT gravity can be expressed \textcolor{black}{in terms of the Hartle-Hawking wavefunction of the disk (in energy basis)} as given in \cite{Saad:2019pqd}. A similar approach can be adopted for the deformed case. The deformed trumpet wavefunction in the length basis is given by, 

\begin{align}
    \begin{split}
\psi^{\text{trumpet}}_{\beta/2,\lambda} (l,b) = \int_0^\infty dE \,\rho_{\lambda}^{\text{trumpet}}(b,{E})\, \psi_{E,\lambda}(\ell)e^{-\beta E/2} \,.
    \end{split}
\end{align}
Now, the propagator in the case of a deformed trumpet can be calculated via the length basis integral decomposition as follows (as described pictorially in Fig.~\eqref{fig12}),
\begin{align} \label{neweq1}
P_{\text{trumpet}}^\lambda(T/2,b,\ell,\ell')=e^{-S_0}\int_{-\infty}^\infty e^{\ell''}dl''\underbrace{\langle \ell,b|\ell''\rangle}_{\textrm{Euclidean part}}\,\,\underbrace{ P_{\chi=1}^\lambda(T/2,\ell'',\ell')}_{\textrm{Lorentzian part}}\,.
\end{align}
One first writes the propagator as,
\begin{align}P_{\text{trumpet}}^\lambda(T/2,b,\ell,\ell')\sim \langle \ell,b|e^{-i\frac{T}{2}H_{\textrm{bulk}}}|\ell'\rangle_{\lambda}\,.\label{5.10m}\end{align}
Now we can readily calculate the matrix element in \eqref{5.10m} as,
\begin{align}
\begin{split}
\langle \ell,b|e^{-i\frac{T}{2}H_{\textrm{bulk}}}|\ell'\rangle_{\lambda}&
=\int_{0}^\infty dE dE' \frac{\langle \ell,b|E'\rangle}{\langle E'|E'\rangle}\langle E'|e^{-iT/2 H_{\textrm{bulk}}}|E\rangle\frac{\langle E|\ell'\rangle}{\langle E|E\rangle}\,,\\&
=\int_0^\infty dE\, \rho_\lambda^{\text{trumpet}}(b,E)e^{-iTE}\psi_{E,\lambda}(\ell)\psi_{E,\lambda}(\ell')\,.\label{4.12m}
\end{split}
\end{align}
We change variables from $E(1-2\lambda E)\to E$. Hence, the Lorentzian part of the emission probability of the baby universe is given by,
\begin{align}
    \langle \ell,b|e^{-i\frac{T}{2}H_{\textrm{bulk}}}|\ell'\rangle_{\lambda}=\int_0^\infty dE\,\frac{\cos\left(b\sqrt{E}\right)}{2\pi\sqrt{E}}e^{-iT E(\lambda)}\psi_{E}(\ell)\psi_{E}(\ell')\,.\label{5.13m}
\end{align}
\textcolor{black}{This integral for non-zero $T$ can be performed numerically, which gives the following behaviour for the probability (after taking the square of the absolute value of the quantity in (\ref{5.13m})) as shown in Fig.~\eqref{fig:2a}. We performed a numerical integration and scanned the value of the integral for different values of $T$ using the inbuilt interpolation function in \textit{Mathematica}.}
\textcolor{black}{
As advocated in \cite{Saad:2019pqd}, for the JT gravity,  Lorentzian evolution is not necessary to compute the baby universe emission probability, as also evident from the above equation (\ref{5.13m}). Even if one sets $T=0$ in (\ref{5.13m}), one still gets a non-vanishing amplitude for the JT gravity. But note that, if one sets $T=0$ in (\ref{5.13m}), all the $\lambda$ dependence vanishes as it appears only in the exponential. So, to investigate the possible effect of the correction due to the deformation parameter, one needs to have a non-vanishing Lorentzian time evolution. }
We found that after switching on the Lorentzian evolution, the baby universe emission amplitude is changed compared to the undeformed theory, as shown in Fig.~\eqref{fig:2a}. We find that, emission rate of baby universes is higher for pure JT gravity than the $T\bar{T}$ deformed theory. \\ 
\textcolor{black}{Now, we intend to compute the non-perturbative (in $\lambda$ as well as in terms of genus sum) aspects of the deformed theory, primarily focusing on the partition functions, density correlators, and possible matrix model. The first task is to compute the partition functions.}
\section{Double Trumpet Partition Function} \label{sec5new}

The double trumpet partition function is obtained by gluing two trumpets with a specific measure that takes care of the relative twist between the two, while joining \cite{Saad:2019lba}. It is usually proportional to the circumference of the geodesic boundary.
Hence, the deformed partition function $Z_{0,2}(\beta_1,\beta_2,\lambda)$ for the double trumpet geometry is obtained by gluing two $T\bar{T}$ deformed trumpets along the common geodesic boundary as,
\textcolor{black}{
\begin{equation}
    Z_{0,2}(\beta_1,\beta_2;\lambda) = \int_0^\infty bdb Z_\lambda^{\text{trumpet}}(b,\beta_1) Z_\lambda^{\text{trumpet}}(b,\beta_2)\label{5.1h}
\end{equation}}
\noindent
\textcolor{black}{To make meaningful comments on a possible matrix-model dual, non-perturbative results are essential. We therefore begin, in the following subsection, by analyzing the non-perturbative partition functions.}
\\\\
\textbf{Non-perturbative computation:} \textcolor{black}{In this subsection, our aim is to compute the non-perturbative density-density correlator. For that, we need the non-perturbative expression for the double trumpet partition function \eqref{5.1h}}. 
As the $ b$-integral is difficult to perform due to the complicated form of the deformed partition function, we write down the series representation of the partition function and, after performing the integral, try to resum the series using the Borel resummation technique.
With a simple change of the variable, one can cast the deformed partition functions in the following way,
\begin{align}
   & Z^{\text{disk}}_{\lambda}=\frac{1}{2\pi^2}\int_0^\infty ds \, s\, \textrm{sinh}(2\pi s)\,e^{-\mathscr{P}(\beta,\lambda,s)}\,,\\&
 Z^{\text{trumpet}}_{\lambda}(\beta,t,b)=\frac{1}{\pi}\int_0^\infty ds \, s\, \textrm{cos}(b\, s)\,e^{-\mathscr{P}(\beta,\lambda,s)} 
\end{align}
where, \begin{align}
    \mathscr{P}(\beta,\lambda,s)=\frac{\beta}{4\lambda}(1-\sqrt{1-8\lambda s^2})\,.
\end{align}
The exponential can be expressed as \cite{Griguolo:2021wgy}\footnote{\textcolor{black}{For $\lambda>0$ the above function has branch point at $s=\frac{1}{\sqrt{8\lambda}}$.}},

\begin{align}
e^{-\mathscr{P}(\beta,\lambda,s)}=e^{-\frac{u\, s^2}{2}}\left(1+\sum_{n=1}^\infty \mathcal{A}_n(s,\beta)\lambda^n\right)\,.
\end{align}
The coefficients can be collectively written in terms of Laugerre polynomials as \cite{Griguolo:2021wgy},
\begin{align}
    \mathcal{A}_n(s,\beta)=-\frac{\beta s^{n+2}}{2^{2n+1}n}L^{n+1}_{n-1}\left(\frac{\beta\, s^2}{2}\right)\,.
\end{align}
Now, using the series representation of the $\sinh$ and $\cos$ functions, we obtain the final form,

\begin{align}
Z_\lambda^{\text{disk}}(\beta,\lambda)=\sum_{n=0}^\infty Z_n^{\text{disk}}(\beta)\,\, \lambda^n\,,\,\,
Z_\lambda^{\text{trumpet}}(\beta,b,\lambda)=\sum_{n=0}^\infty Z_n^{\text{trumpet}}(\beta,b)\,\, \lambda^n
\end{align}
where,
\begin{align}
\begin{split}
   & Z_n^{\text{disk}}(\beta)=\frac{(2\beta)^{-n}}{n!\sqrt{2\pi^3\beta^3}}\Gamma(n-\frac{3}{2})\Gamma(n+\frac{5}{2}) \,_1F_1\Bigg(n+\frac{5}{2};\frac{5}{2}-n;\frac{2\pi^2}{\beta}\Bigg)\,,\\&
   Z_n^{\text{trumpet}}(\beta,b)=-\frac{(2\beta)^{-n}}{n!\sqrt{2\pi^3\beta}}\Gamma(n-\frac{1}{2})\Gamma(n+\frac{3}{2}) \,_1F_1\Bigg(n+\frac{3}{2};\frac{3}{2}-n;-\frac{b^2}{2\beta}\Bigg)\,.
   \end{split}
\end{align}\\

\noindent
\textcolor{black}{After Borel re-summation, we have the following closed form expression of the disk and trumpet partition functions, respectively.}
\textcolor{black}{\begin{align}
    \begin{split}
    \label{5.9u}
&Z_\lambda^{\text{disk}}(\beta) = \frac{\beta}{4\sqrt{2\lambda}}\frac{e^{-\frac{\beta}{4\lambda}}}{\beta^2 + 8\pi^2\lambda}I_2\left(\frac{\sqrt{\beta^2 + 8\pi^2\lambda}}{4\lambda}\right)\,,\\&
Z_\lambda^{\text{trumpet}}(b,\beta) = \frac{\beta}{2\sqrt{2\lambda}} \frac{e^{-\frac{\beta}{4\lambda}}}{\sqrt{\beta^2 - 2b^2\lambda}}I_1\left(\frac{\sqrt{\beta^2 - 2b^2\lambda}}{4\lambda}\right)\,.
    \end{split}
\end{align}}
\textcolor{black}{Here $I_n$ is the modified Bessel function of the first kind.}
Furthermore, borrowing the result from \cite{Griguolo:2021wgy}, we directly use the non-perturbative form of the double trumpet partition function in \eqref{5.9u} and glue two trumpets to form the Euclidean two-boundary wormhole whose partition function can be expressed as, 

\begin{align}
    Z_{0,2}(\beta_1,\beta_2;\lambda)= \frac{\beta_1\beta_2 e^{-(\beta_1+\beta_2)/\lambda}}{\lambda(\beta_1^2-\beta_2^2)}\Bigg[\beta_1 I_0\left(\frac{\beta_2}{\lambda}\right)I_1\left(\frac{\beta_1}{\lambda}\right)-\beta_2 I_0\left(\frac{\beta_1}{\lambda}\right)I_1\left(\frac{\beta_2}{\lambda}\right)\Bigg]\,. \label{5.10i}
\end{align}
\noindent

This will play a crucial role in the computation of the density-density correlator, which will be needed for the computation of the length of the Einstein-Rosen Bridge (ERB). Before we delve into that computation, we will need a few more ingredients as well as input from the matrix model. We now discuss them. \\
\section{$T\bar{T}$ deformed Matrix model} \label{sec5}
\noindent
\textbf{Matrix integral and resolvents:}
Taking a cue from \cite{Saad:2019lba}, we briefly discuss the connection with the matrix integral and the resolvents, which are well studied in Random matrix theories (RMTs).  In RMT, the partition function is written using a matrix integral in the following way \cite{Eynard:2015aea},
\begin{align}
    \mathcal{Z}=\int dH e^{-N Tr V(H)}\,,\hspace{2 cm} H=N\times N \textrm{ Hermitian Matrix}\,.
\end{align}
\noindent
The observables are $\textrm{Tr}\, e^{-\beta H}$ and the expectation value of such observables is given by,

$$\langle Z(\beta_1)\cdots Z(\beta_n)\rangle= \frac{1}{\mathcal{Z}}\int dH \,e^{-N Tr V(H)}Z(\beta_1)\cdots Z(\beta_n)\,.$$
\noindent
In this context, {resolvents} play an important role.  These are defined by,

\begin{align}
    R(E)=\textrm{Tr}\,\frac{1}{E-H}=\sum_{i=1}^N\frac{1}{E-\lambda_i}\,.
    \end{align}
Here, $E$ denotes an arbitrary complex number.  For a fixed Hermitian matrix $(H)$, this sum over poles corresponding to the eigenvalues of $H$ is smeared into branch cuts after taking averages \cite{Saad:2019lba}. 
The correlation function of the resolvents admits a $1/N$ expansion of the following nature,

\begin{align}
    \langle R(E_1)\cdots R(E_n)\rangle\sim \sum_{g=0}^\infty \frac{R_{g,n}(E_1,\cdots,E_n)}{N^{2g-n-2}}\,.
\end{align}
 In large-$N$ limit, one can show that \cite{Saad:2019lba},
 \begin{align}
   R (E+i\epsilon)+R(E-i\epsilon)=V'(E)
\end{align}
\noindent
where $V(E)$ is the matrix potential. Resolvents have a nice connection to the partition functions.  They are, in fact, the Laplace transform of the partition function,
\begin{align}
R(E)=-\int_0^\infty d\beta \,e^{\beta E}Z(\beta)\,.\label{double_trumpet}
\end{align}
This integral makes sense for $E$ less than the ground state energy.  We can now use \eqref{double_trumpet} to compute the following correlator of resolvent functions, which is defined as,
\begin{align}
    \begin{split}
        R_{0,2}(E_1,E_2;\lambda)& = \int_0^\infty d\beta_1\,d\beta_2\,  Z_{0,2}(\beta_1,\beta_2;\lambda)\, e^{\beta(E_1 + E_2)}\,.
    \end{split}
\end{align}
Then, using (\ref{5.10i}), we get, 
\textcolor{black}{\begin{align}
\begin{split}
   & R_{0,2}(E_1,E_2;\lambda)=\frac{64
\lambda^2(1-8\lambda E_1/2)(1-8\lambda E_2/2)\left(8\lambda(E_1^2+E_2^2)/4-E_1-E_2\right)}{4\left[(1-8\lambda E_1/2)^2-(1-8\lambda E_2/2)^2\right]^2\sqrt{-E_1(1-\frac{8\lambda E_1}{4})}\sqrt{-E_2(1-\frac{8\lambda E_2}{4})}}\\&\hspace{7 cm}-\frac{8\lambda^2\left[(1-8\lambda E_1/2)^2+(1-8\lambda E_2/2)^2\right]}{4\left[(1-8\lambda E_1/2)^2-(1-8\lambda E_2/2)^2\right]^2}\,.\label{3.31t}
    \end{split}
\end{align}}
Keeping terms upto first order in $\lambda$ we get, 
 \begin{align}
     \begin{split}
         R_{0,2}(E_1,E_2;\lambda) = \frac{1}{4\sqrt{-E_1}\sqrt{-E_2}(\sqrt{-E_1}+\sqrt{-E_2})^2} + \frac{\lambda}{4\sqrt{-E_1}\sqrt{-E_2}}+\mathcal{O}(\lambda^2)\,. \label{3.31ta}
     \end{split}
 \end{align}
Here, the $\lambda$ independent term agrees with the correlator of resolvent functions for pure JT gravity \cite{Saad:2019lba}. 
It is useful to express this in terms of the variable $z$ where $z_i = \sqrt{-E_i}$
 \begin{equation}
     R_{0,2}(z_1,z_2;\lambda) = \frac{1}{4z_1z_2(z_1+z_2)^2} + \frac{\lambda}{4z_1z_2}\,\,.\label{3.47m}
 \end{equation}
Now this $R_{0,2}(z_1,z_2;\lambda)$ will be used to compute the density-density correlation.  Before doing that, we first discuss the nature of the  $T\bar{T}$ deformed matrix model, as we will borrow several technologies from the matrix model side to compute the density-density correlator.  The above expression for $R_{0,2}(z_1,z_2;\lambda)$ will be very useful to arrive at the conclusion about the branch cut structure of the underlying matrix model i.e it is a single-cut or multi-cut model. 
\subsection*{Comments on the cut structure of the underlying matrix model}
Before proceeding further, we need to discuss in detail the underlying cut-structure of the matrix model.  We take a similar route to what has been done for the case of pure JT gravity \cite{Saad:2019lba}.  We perform the following steps.
\begin{figure}
    \centering
\includegraphics[width=0.4\linewidth]{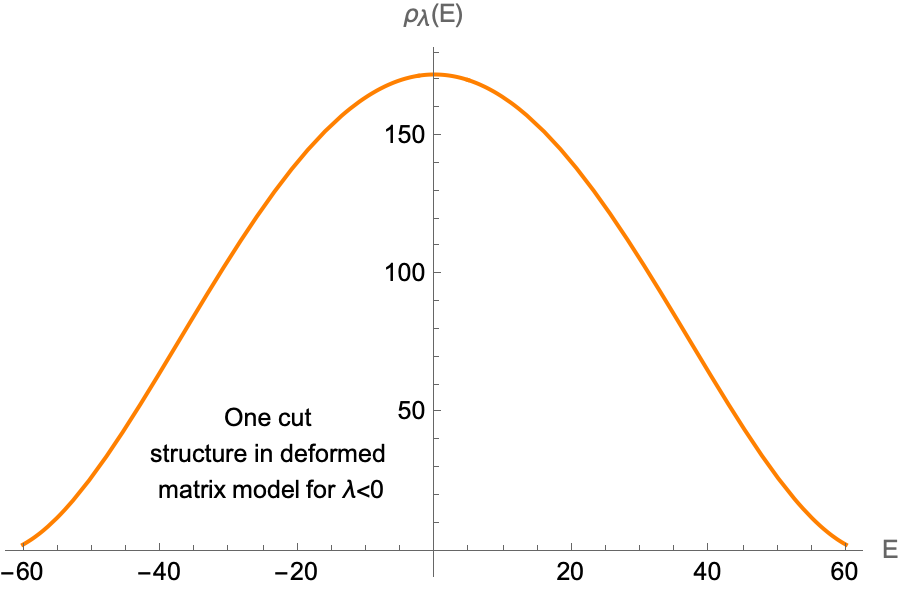}\hspace{1.2 cm}
\includegraphics[width=0.4\linewidth]{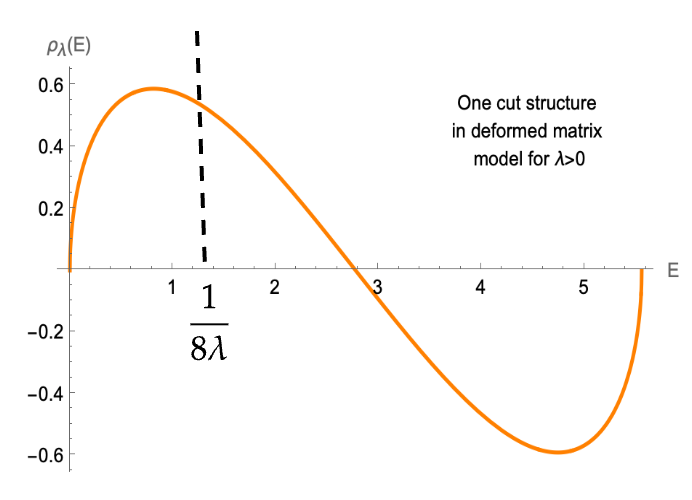}
    \caption{Schematic diagram showing that JT gravity with $T\bar{T}$ deformation shows one-cut structure for $\lambda<0$ (left figure). It also shows a one-cut structure for $\lambda>0$ (right figure) in the sense that $R_{0,2}$ takes the universal form at the two ends of the cut i.e at $(0,\frac{1}{2\lambda})\,.$ But we can also see that, for $\lambda>0$ the density becomes negative at $E=\frac{1}{4\lambda}$, making this one-cut structure of the matrix model unfeasible. The left figure is after scaling the density as mentioned in \eqref{5.11u},  where we have chosen $a=60$ while plotting the figure for $\lambda<0$, and we found the plot has support between $(-a,a)\,.$ The right figure is without scaling, and the dotted line represents the cut-off that needs to be imposed for $\lambda >0$ to keep the spectrum real (refer to (\ref{3.15k})).}
    \label{fig:4}
\end{figure}
\begin{itemize}
\item For a generic sign of the deformation parameter, if we first perform the \textcolor{black}{following replacement, which is usually done to extract physics near the cut of the finite-cut matrix model,}
\begin{align}
    \sqrt{E}\to \sqrt{\frac{-E^2+a^2}{2a}}\,.\label{5.11u}
\end{align}
\textcolor{black}{The density becomes the following,}
\begin{align}
    \rho_{\lambda}(E)^{\text{rescaled}}= \frac{\Big(1-4\lambda {\frac{a^2-E^2}{2a}}\Big)}{4\pi^2}\sinh\Bigg(2\pi/\sqrt[4]{2} \sqrt{(E\pm a)(E-E_{+})(E-E_{-})}\,\Bigg) \label{resc}
\end{align}
where, $$E_{\pm}=\pm\frac{\sqrt{a} \sqrt{a \lambda +1}}{\sqrt{\lambda }}\,.$$ 
At this point, it is tempting to conjecture that this matrix model possesses a double-cut structure for both the good and bad signs of $\lambda$ under this scaling, but things are more subtle as we will explain shortly.
\item Also, for a multi-cut ($s$-cut) matrix model, we have the following expression for the planar one-loop resolvent \cite{Marino:2015yie},
\begin{align}
    R_{0,1}^s(E):=\frac{1}{2}\oint_C\frac{d\omega}{E-\omega}\sqrt{\prod_{i=1}^{2s}\frac{E-x_i}{\omega-x_i}}
\end{align}
where $x_i$ denotes the $2s$ edges for a $s$-cut matrix model.
From this, one can read off the density for multicut models as follows,
\begin{align}
    \rho(E)&=\frac{1}{2\pi i}\lim_{\epsilon\to 0}(R_{0,1}(E-i\epsilon)-R_{0,1}(E+i\epsilon))\,,\\&
    =\frac{1}{2\pi}|M(E)|\sqrt{-\prod_{i=1}^{2s}(E-x_i)}\,.
\end{align}
For specific cases such as double-cut matrix models, we can express the universal form of planar two-loop resolvent as follows \cite{Akemann:1995im},
\begin{align}
\begin{split}
    R_{0,2}^{s=2}(E_1,E_2):=\frac{1}{4(E_1-E_2)^2}&\Bigg(\sqrt{\frac{(E_1-x_1)(E_1-x_4)(E_2-x_2)(E_2-x_3)}{(E_1-x_2)(E_1-x_3)(E_2-x_1)(E_2-x_2)}}\\&\hspace{1 cm}+\sqrt{\frac{(E_1-x_2)(E_1-x_3)(E_2-x_1)(E_2-x_4)}{(E_1-x_1)(E_1-x_4)(E_2-x_2)(E_2-x_3)}}\Bigg)\\&
    +\frac{1}{4}\frac{1}{\sqrt{\prod_{j=1}^{4}(E_1-x_j)(E_2-x_j)}}\frac{E(\mathscr{K})}{K(\mathscr{K})}f(\{x_i\})-\frac{1}{2(E_1-E_2)^2}
\end{split}
\end{align}
where,
\begin{align}
\begin{split}
f(\{x_i\}):&=\frac{x_2x_3x_4(x_2-x_4)}{(x_1-x_4)(x_2-x_1)}+\frac{x_1x_3x_4(x_1-x_3)}{(x_1-x_2)(x_2-x_3)}\\&\hspace{1 cm}+\frac{x_1x_2x_4(x_4-x_2)}{(x_2-x_3)(x_3-x_4)}+\frac{x_1x_2x_3(x_1-x_3)}{(x_1-x_4)(x_3-x_4)}
\end{split}
\end{align}
and
\begin{align}
    \begin{split}
&K(\mathscr{K})=\int_{0}^1 dt \frac{1}{\sqrt{(1-t^2)(1-\mathscr{K}^2t^2)}},\hspace{1 cm}
E(\mathscr{K}):=\int_{0}^{1}dt\, \sqrt{\frac{1-\mathscr{K}^2t^2}{1-t^2}}\,.
    \end{split}
\end{align}
Now it is evident from our previous expressions (e.g (\ref{3.31t})) that the $T\bar{T}$-deformed matrix model, in particular, does not exhibit the elliptic integrals that are typically expected from the universal behavior of a double-cut matrix model.  This absence indicates that the $T\bar{T}$-deformed matrix model perhaps remains in the one-cut phase.

 \item Now we observe that, for the good sign of $\lambda$ i.e when $\lambda < 0$, if one plots the density, it shows a single cut structure (refer to the left panel of the Fig.~(\ref{fig:4}))\,. Furthermore, to solidify our claim about the single-cut nature of the matrix model (after scaling of $E$ as mentioned in \eqref{5.11u}) the planar two-loop resolvent in the \textbf{coincident limit} (obtained from \eqref{3.31t}) is given by,
    \begin{align}
        R_{0,2}^{\text{scaled}}(E,E)=\frac{a^4 \left({8 \lambda  \left(a^2-\text{E}^2\right) \left(a^2 \lambda +a+\text{E}^2 (-\lambda )\right)}+a^2\right)}{4 \left(\text{E}^2-a^2\right)^2 \left(a^2 \lambda +a+\text{E}^2 (-\lambda )\right)^2 \left(2 a^2 \lambda +a-2 \text{E}^2 \lambda \right)^2}\,. \label{newquation}
    \end{align}
 This, near the two edges of the cut i.e $(-a,a)$ takes the following form (after  expanding around these two values of $E$ and keeping only the leading order term), 
\begin{align}
R_{0,2}^{\textrm{scaled}}(E,E) =
\begin{cases}
\displaystyle \frac{1}{16 (E+a)^2}\,, \\
\displaystyle \frac{1}{16\left(E -a\right)^2}\,.\label{6.20inew}
\end{cases}
\end{align} 

Now it is well known that the one-cut matrix models known to possess a specific universal form for $R_{0,2}(E_1,E_2)$ near the two edges of the cut, with $E_1\to E_2$, and is given by \cite{Akemann:1995im},
\begin{align}
    R_{0,2}(E_1,E_1)=\frac{C}{16(E_1-z_i)^2}\label{6.20i}
\end{align}
where $z_i$ denote the edge points.
Now it is evident by comparing (\ref{6.20inew}) and (\ref{6.20i}) that there is a universal behavior.  Importantly, one can also check that, if we expand (\ref{newquation}) around the other points $E_{\pm}$ as mentioned in (\ref{resc}), we don't observe any universal behaviour of  $R_{0,2}(E_1,E_1)$ like what has been mentioned in (\ref{6.20i}).  So the apparent double cut structure as mentioned below (\ref{6.20i}) is not there and the model possesses only a single-cut structure.  We now perform the double scaling \cite{Saad:2019lba} by replacing  $E\to E-a$ in (\ref{resc}) and taking  $a\to \infty\,$. This was done previously for JT gravity as well to show that the underlying matrix model is a leading approximation of the single-cut matrix model \cite{Saad:2019lba}.  Similarly, for our case, after the double scaling limit, $\rho_{\lambda}(E)$ goes back to (\ref{3.16y}) at the leading order and we can also conclude that, for $\lambda < 0,$ our density corresponds to that of the density of a single cut matrix model at the leading order (in the expansion of large $a$).  It is worth noting that, in the large-$N$ limit, holographic interpretations are generally associated only with one-cut matrix models.  \textit{Hence, keeping in mind this, we will henceforth restrict our calculation to the good sign of $\lambda$.}

\item  Before we end, we make comments about the $\lambda >0\,.$ Note that, even if we do not perform the scaling as mentioned in (\ref{5.11u}), we can see that for  $\lambda < 0 $ from (\ref{3.16y}), there will be a branch cut that will fall between $0$ and $-\frac{1}{2 |\lambda|}\,.$ But this is problematic in that sense, the density will become imaginary in this range.  Hence, the scaling (\ref{5.11u}) and the subsequent analysis, as discussed above, is extremely crucial to comment on whether the underlying model is one-cut or not.  In this spirit, we also note that, for $\lambda >0 $ (bad sign), the branch cut will run between ($0,\frac{1}{2\lambda}$))\,. So one may be tempted to conclude that, for $\lambda >0$ the underlying model is a one-cut one.  In fact, one can also check the universality of the $R_{0,2}$ around the edges as discussed for $\lambda <0\,.$ One will find that it displays universal behaviour  in the coincident limit i.e $E_1 \rightarrow E_2$ at the two edges $E_2=0$ and $E_2=\frac{1}{2\lambda}$  (again obtained from \eqref{3.31t})\,.
\begin{align}
R_{0,2}(E_2,E_2) =
\begin{cases}
\displaystyle \frac{1}{16 E_2^2}, \\
\displaystyle \frac{1}{16\left(E_2 - \frac{1}{2\lambda}\right)^2}\,.
\end{cases}
\end{align}

But one has to be careful here, as at $E=\frac{1}{4\lambda}$ there will be a node, where the density vanishes as shown in Fig.~(\ref{fig:4}).  We also know that, beyond $E=\frac{1}{8\lambda}$ the spectrum becomes imaginary. So, even with this universality at the endpoints, the matrix model density for $\lambda >0$ is unfeasible.  Furthermore, even under the double scaling as discussed earlier, we can easily show that for $\lambda >0,$ there is no one cut structure like what has been shown for $\lambda <0$ in the left panel of Fig.~(\ref{fig:4}).  So we only have a one-cut matrix model for $\lambda <0$, and the density that we work with is the leading order approximation of that under the double scaling limit.\\
\end{itemize}
\textcolor{black}{
To summarize, the candidate $T\bar{T}$-deformed matrix model exhibits an apparent one-cut behavior\footnote{$T\bar{T}$-deformed matrix models do not admit a consistent single-cut solution without an appropriate scaling.} upon implementing the scaling of $E$ as described in \eqref{5.11u}. In this regime, the planar two-loop resolvent near the edges displays universal behavior for the bad sign of the deformation parameter.
Although the resolvent exhibits  universality near the edges i.e at $(0, \frac{1}{2\lambda})$ of the cut, we must restrict the spectrum to $E \leq \frac{1}{8\lambda}$ (in particular, $< \frac{1}{2\lambda}$), since beyond this point the spectrum becomes complex.  However, this restriction effectively destroys the one-cut structure.
 To obtain a consistent one-cut matrix model structure, we instead perform an appropriate scaling of $E$.  We then observe that for $\lambda < 0$, the seemingly double-cut structure exhibits universality only at the two endpoints, $(-a, a)$.  The absence of universal behavior at the remaining endpoints indicates that the model effectively reduces to a one-cut structure for $\lambda < 0$ also shown in figure~\eqref{fig:4}.
}
\vspace{0.5 cm}
\noindent

\definecolor{LightGray}{rgb}{0.9,0.9,0.9}
\begin{table}[ht!]
\centering
\begin{tabularx}{\textwidth}{|X|X|}
\hline
\textbf{Theory + deformation type} & \textbf{Matrix-model type} \\
\hline
\rowcolor{LightGray}
Pure JT gravity & Comes from one-cut model in the double scaling limit. \\
JT gravity + $T\bar{T}$ deformation (+ve sign of $\lambda$) 
&No feasible one-cut structure after scaling.  Without any scaling, the density becomes negative after $E>\frac{1}{4\lambda}$.\\
\rowcolor{LightGray}
JT gravity + $T\bar{T}$ deformation (-ve sign of $\lambda$) 
& Comes from one-cut matrix model via scaling,\textit{ with density becoming positive on the cut support}.\\
\hline
\end{tabularx}
\caption{Matrix model type for different signs of the deformation parameter.}
\label{tab1}
\end{table}


\subsection{ $T\bar{T}$ deformed matrix potential and density correlator}
\begin{figure} [htb!]
\centering
\scalebox{0.6}{\includegraphics{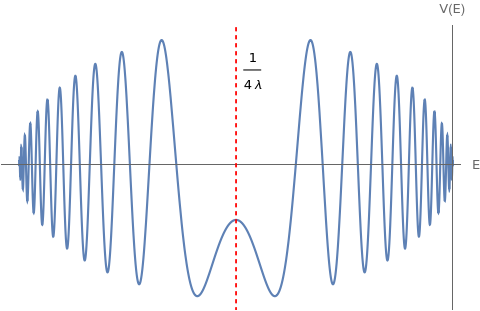}}
\caption{Effective matrix potential for $T\bar{T}$ deformed JT gravity with a saddle point at $\frac{1}{4\lambda}$ marked by the dashed line.}\label{fig2}
\end{figure}

Now start computing the density-density correlator.  We first find the deformed spectral curve and calculate the deformed matrix potential.  The spectral curve of the matrix integral can be written as  \cite{Saad:2019lba},
\begin{align}
    \begin{split}
        y(E) = -i\pi\rho^{\text{disk}}_\lambda(E)
    \end{split}
\end{align}
where $\rho^{\text{disk}}_\lambda(E)$ is the deformed density of states as shown in  \eqref{3.16y}, for which the leading order contribution comes from the disk topology.  Now, defining the following variable,
\begin{equation}
    z^2 = -E
\end{equation}
we get
\begin{align}
    \begin{split}
        y(z) = (1+4\lambda z^2)\frac{\sin(2\pi z\sqrt{1+2\lambda z^2})}{4\pi}\,.
        \label{deformed_spectral_curve}
    \end{split}
\end{align}
Given the discussion in Sec~(\ref{sec5}), we focus only on the good sign of the deformation parameter.  Since it is better to work with determinants instead of resolvents, especially if one wants to consider the non-perturbative effects, we define the following quantity, which we shall refer to as the `determinant.'
\begin{equation}
    \psi(E) \equiv \text{det}(E-H)e^{-\frac{L}{2}V(E)}\,. \label{new1}
\end{equation}
To compute the expectation value $\langle \psi \rangle$ we need to consider an integral over $H$ with a weighting that depends on the deformed matrix potential $V(H,\lambda)$
\begin{equation}
    \langle \text{det}(E-H) \rangle = \frac{1}{\mathcal{Z}} \int dH\, \text{det}(E-H) e^{-L \text{Tr}\, V(H,\,\lambda)} 
\end{equation}
where the normalization $\mathcal{Z}$ is given by, 
\begin{equation}
    \mathcal{Z} = \int dH\, e^{-L \text{Tr}\, V(H,\,\lambda)} \,.
\end{equation}
The deformed matrix potential can be calculated using the deformed spectral curve as follows
\begin{align}
    \begin{split}
        V_{\text{eff}}(E) &= 2 e^{S_0} \int_0^{-E} dx\, y(\sqrt{x})\,,\\
        &= \frac{e^{S_0}}{2\pi} \int_0^{-E} dx\, (1+4\lambda x){\sin\big(2\pi \sqrt{x(1+2\lambda x)}\big)}\,.
    \end{split}
\end{align}
Changing the variable of integration to $k(x) = \sqrt{x(1+2\lambda x)}$ we have
\begin{align}
    \begin{split}
        V_{\text{eff}}(E) &= \frac{e^{S_0}}{\pi} \int_0^{k(-E)} kdk \sin(2\pi k)
        \,,\\
        &= \frac{e^{S_0}}{4\pi^3} \big( \sin[2\pi k(-E)] - 2\pi k(-E) \cos[2\pi k(-E)] \big)\,.\\
       \label{5.8u}
    \end{split}
\end{align}
A plot for this effective potential is given in Fig.~\eqref{fig2}.


Before proceeding further,
we comment on the nature of $ V_{\text{eff}}(E)\,.$
For pure JT gravity, the potential is given by \begin{align}
    V_{\text{eff}}(E)=\frac{e^{S_0}}{4\pi^3}\Bigg[\textrm{sin}(2\pi \sqrt{-E})-2\pi \sqrt{-E}\,\textrm{cos}(2\pi \sqrt{-E})\Bigg]\,, \hspace{2 cm} E<0\,.
\end{align}
As mentioned in \cite{Saad:2019lba}, this potential is non-perturbatively unstable.  After $E=-1/4$, it oscillates rapidly, and it reaches a local maximum at this specific point.  One way to get rid of this oscillation is using the {`one-eigenvalue instantons'} \cite{Saad:2019lba}.  They are sometimes called as {\textit{ ZZ branes}}.  However, they are manifestly different from the FZZT branes, which one can also insert to control these oscillations \cite{Saad:2019lba}.
But for our case, the $T\bar{T}$ deformed matrix potential (\ref{5.8u}) already stops its erratic oscillation near $E\sim\frac{1}{4\lambda}$. It seems that ZZ brane states are natural here.
Now, to compute $\langle \psi (E) \rangle$ from (\ref{new1}) we use the following identity
\begin{equation}
    \text{det}(E-H) = \text{exp}\big(\text{Tr}\log(E-H)\big)\,.
\end{equation}
We can then express the expectation value as follows
\begin{align}
    \begin{split}
       \langle  \text{det}(E-H) \rangle =  \text{exp}\bigg(\big\langle\text{Tr}\log(E-H)\big\rangle + \frac{1}{2}\big\langle\text{Tr}\log(E-H)\text{Tr}\log(E-H)\big\rangle_{\text{conn}} + \cdots\bigg)\,.
       \label{expectation_of_det}
    \end{split}
\end{align}
The first term in the exponential contributes to the disk amplitude, and the second term in the exponential contributes to the cylinder amplitude, which we will evaluate below eventually.  Following \cite{Saad:2019lba}, disk amplitude can be evaluated by the planar one-loop resolvent as follows,
\begin{equation}
\left\langle \textrm{Tr} \log(E - H) \right\rangle_{g=0}
= e^{S_0} \int^E dE'\, R_{0,1}(E') \,.
\label{eq:genus0_resolvent}
\end{equation}
Now, combining with the exponential of the potential $e^{-\frac{L}{2} V(E)}$ mentioned in (\ref{new1}) and using the following relation from the matrix model,
\begin{equation}
e^{S_0} y(E) = e^{S_0} R_{0,1}(E) - \frac{L}{2} V'(E) \, ,
\end{equation}
we obtain the expectation value of the `determinant' defined in \eqref{new1}, in the leading order as follows,
\begin{equation}
\left\langle \psi(E) \right\rangle
= e^{-\frac{L}{2} V(E)} \left\langle \det(E - H) \right\rangle
\;\sim\;
\exp\left( \underbrace{e^{S_0} \int^E dE'\, y(E')}_{\text{Disk}(z)} \right) \, ,
\label{eq:leading_psi}
\end{equation}
Hence, we compute this Disk$(z)$ amplitude for the deformed spectral curve now.  The disk amplitude as a function of $z$ is given by \cite{Saad:2019lba},
\begin{align}
    \begin{split}
        \text{Disk}(z) \equiv e^{S_0}\int_0^z y(z')(-2z'dz')\,.
    \end{split}
\end{align}
Using the spectral curve for the deformed theory as given in \eqref{deformed_spectral_curve}, we get,
\begin{align}
    \begin{split}
        \text{Disk}(z) = \frac{e^{S_0}}{4\pi}\int_0^z (1+4\lambda z'^2)\sin(2\pi z'\sqrt{1+2\lambda z'^2})(-2z'dz')\,.
    \end{split}
\end{align}
To compute the integral again, we change the integration variable to $k(z') = \sqrt{z'(1+2\lambda z'^2)}\,.$
\begin{align}
    \begin{split}
        \text{Disk}(z) &= -\frac{e^{S_0}}{2\pi} \int_0^{k(z)} \sin(2\pi k) kdk\,,\\
        &= - \frac{\sin(2\pi k(z))- 2\pi k(z)\cos(2\pi k(z))}{8\pi^3}e^{S_0}\,.
    \end{split}
\end{align}
Now, the second connected term in the exponential (which we define as cylinder amplitude) is given by \cite{Saad:2019pqd},
\begin{align}
    \begin{split}
        \text{Cyl}(z_1,z_2;\lambda) \equiv \int_\infty^{z_1} (-2z'_1dz'_1)\int_\infty^{z_2} (-2z'_2dz'_2) R_{0,2}(z'_1,z'_2;\lambda)\,.\label{5.15u}
    \end{split}
\end{align}
Using the correlator of resolvents as shown in equation \eqref{3.47m} , we have
\begin{align}
    \begin{split}
        \text{Cyl}(z_1,z_2;\lambda) &= \int_\infty^{z_1} (-2z'_1dz'_1)\int_\infty^{z_2} (-2z'_2dz'_2) \left(\frac{1}{4z'_1z'_2(z'_1+z'_2)^2} + \frac{\lambda}{4z'_1z'_2}\right)\,,\\
        &= -\log(z_1 + z_2) + \lambda z_1 z_2 + \textcolor{black}{\text{divergent term}\,.}\, \label{new2}
    \end{split}
\end{align}
Finally inserting all these in \eqref{expectation_of_det}, we get (after discarding the divergent term from cylinder amplitude), 
\begin{align}
    \begin{split}
        \Psi (z) &= \text{exp}\bigg[\text{Disk}(z) + \frac{1}{2}\text{Cyl}(z,z;\lambda)\bigg]\,,\\
        &= \frac{1}{\sqrt{2z}}\text{exp}\Big[\text{Disk}(z) + \frac{\lambda z^2}{2}\Big]\,.
    \end{split}
\end{align}
\newpage

\noindent
\textbf{Density correlator:}\\

Finally, the density pair correlation function $\langle \rho(E_1)\rho(E_2)\rangle$ for $|E_1 - E_2|\ll 1$ and $E_1,E_2 > 0$ can be computed from the resolvent pair correlator. \textcolor{black}{ In the classically allowed region, the prescription is to sum over both branches for the `determinant', while for the inverse-determinant associated with $R_\pm$, one retains only the branch,
$z = e^{\mp i \frac{\pi}{2} \sqrt{E}} \, .$ Now the two-point correlator of resolvent is given by \cite{Saad:2019lba},
\begin{equation}
    \big\langle R^\pm(E_1) R^\pm(E_2)\big\rangle  = \partial_{E_1}\partial_{E_2} \big\langle \underbrace{\psi(E_1) \psi(E_2)\tilde\psi(E_1)_{\pm} \tilde\psi(E_2)_{\pm}}_{\Psi(z_1,z_2;z_3,z_4)}\big\rangle \big\vert_{E_1 = E_3, E_2 = E_4},\quad \tilde\psi(E)=\frac{e^{\frac{L}{2}V(E)}}{\det(E-H)}\,.\label{6.41w}
\end{equation}}
The density pair correlator can then be found using \cite{Marino:2015yie},
\begin{equation}
    (-2\pi i)^2 \langle \rho\rho \rangle = \langle R^+R^+ \rangle + \langle R^-R^- \rangle - \langle R^+R^- \rangle - \langle R^-R^+ \,\rangle\,.\label{6.42r}
\end{equation}
We restrict our computation to the terms that are singular as $E_1 \rightarrow E_2$.  The one-loop function is given in terms of the disks and cylinders as,\footnote{The $\Psi (z_1,z_2;z_3,z_4)$ notation is used to connect it with the disk and cylinder picture.  The four insertions i.e $z_1,z_2,z_3,z_4$ are dependent on the resolvent arguments.  }\\
\begin{align}
    \begin{split}
        \Psi (z_1,z_2;z_3,z_4) = \text{exp}\bigg[&\text{Disk}(z_1)+\text{Disk}(z_2)-\text{Disk}(z_3)-\text{Disk}(z_4)+ C(z_1,z_3)+C(z_1,z_4)\\&+C(z_2,z_3)+C(z_2,z_4)-C(z_1,z_2)-C(z_3,z_4)\bigg],
    \end{split}
\end{align}
where, 
\begin{align}
    \begin{split}
        C(z,z') &= \frac{1}{2}\big( \text{Cyl}(z,z,\lambda)+\text{Cyl}(z',z',\lambda)\big) - \text{Cyl}(z,z',\lambda)\,,\\
        &= \log(\frac{z+z'}{2\sqrt{zz'}}) + \frac{\lambda}{2}(z-z')^2\,.
    \end{split}
\end{align}
We have used (\ref{new2}).  Explicitly $\Psi(z_1,z_2;z_3,z_4) $ is then given as,
\begin{align}
    \begin{split}
        \Psi(z_1,z_2;z_3,z_4) = \frac{(z_1+z_3)(z_1+z_4)(z_2+z_3)(z_2+z_4)}{4\sqrt{z_1z_2z_3z_4}(z_1+z_2)(z_3+z_4)} \text{exp}\bigg[\text{Disk}(z_1)+\text{Disk}(z_2)-\text{Disk}(z_3)-\text{Disk}(z_4)\\+ \frac{\lambda}{2}(z_1^2+z_2^2+z_3^2+z_4^2) -\lambda(z_1+z_2)(z_3+z_4)+\lambda(z_1z_2+z_3z_4)\bigg]\,.\label{6.45r}
    \end{split}
\end{align}

The deformed two-point density correlator can then be written following the computations for pure JT gravity as done in Appendix (2) of \cite{Saad:2019lba} \footnote{To compute it non-perturbatively, one should use the non-perturbative form of the resolvents.}.  \textcolor{black}{Now, one can compute \eqref{6.42r} plugging the expression obtained in  \eqref{6.45r} after performing the derivative with respect to the energies as defined in \eqref{6.41w} as follows,  }\\

\hfsetfillcolor{gray!8}
\hfsetbordercolor{white}
\begin{align}\begin{split}
\tikzmarkin[disable rounded corners=true]{jol}(1.8,-1.08)(-0.8,1.2)
   \hspace{-0.7 cm} 
        \big\langle \rho(E_1)\rho(E_2) \big\rangle \approx&\,\, e^{2S_0}\underbrace{\rho_\lambda^{\text{disk}}(E_1)\rho_\lambda^{\text{disk}}(E_2)}_{\text{\textcolor{black}{Semiclassical disconnected piece}}} + e^{S_0} \underbrace{\rho_\lambda^{\text{disk}} \delta(E_1-E_2)}_{\text{Contact term}} \,\\&-\underbrace{\frac{1}{2\pi^2(E_1-E_2)^2}
        + \frac{e^{-2\lambda(\sqrt{E_1}-\sqrt{E_2})^2}}{2\pi^2(E_1-E_2)^2} \cos[2\pi e^{S_0}\rho_\lambda^{\text{disk}}(E_2)(E_1-E_2)]}_{\text{\textcolor{black}{Modified Non-perturbative contribution at $\mathcal{O}(\lambda)$}}}\,.\label{5.23c}
        \tikzmarkend{jol}
    \end{split}
\end{align}\\
This is one of the main ingredients required for our subsequent study of the growth of the length of ERB. 
One can proceed with non-perturbative form of $R^{\lambda}_{0,2}$ and then calculate the two-point density correlator by finding {Cyl\,($z_1,z_2;\lambda$)} function analogous to (\ref{5.15u}) by not expanding it in terms of $\lambda\,.$ \textcolor{black}{It will be nice to obtain a closed form expression for the density correlator in the deformed matrix model non-perturbatively, though we will be only interested in performing calculations perturbatively in $\lambda$.}\\
Next we proceed to calculate the \textcolor{black}{deformed matrix elements of the two-point function of light insertions at the boundaries of the disc} for $T\bar{T}$ deformed case.

\subsection{Deformed matrix elements}
\noindent
We now have all the ingredients needed to calculate the deformed matrix elements corresponding to the non-spinning operator insertion at the boundary. Following \cite{Saad:2019pqd}, one can identify the deformed matrix element. \textcolor{black}{For the undeformed case with operator insertions of weight $\Delta$ the matrix element is given by,}  
\begin{align}
\begin{split}
  \int_{-\infty}^{\infty}d\ell\, \psi_{E}(\ell)\psi_{E'}(\ell)\,e^{-\Delta \ell}=&\frac{|\Gamma(\Delta+(is_1+is_2))\Gamma(\Delta+(is_1-is_2))|^2}{2^{2\Delta+1}\Gamma(2\Delta)}, \hspace{0.6 cm} \text{with } s_1=\sqrt{E} , s_{2}=\sqrt{E'}\\&
  =\mathcal{M}_{\Delta}(E,E')\,. \label{new3}
\end{split} 
\end{align}
\textit{\textbf{Deformed matrix elements:}}
In undeformed JT gravity, the length of the geodesic connecting the two boundary operators requires regularisation through the introduction of a small UV cutoff $\epsilon$. By contrast, the $T\bar T$ deformation provides a natural cutoff for the boundary, set by the deformation parameter itself. Denoting the corresponding cutoff radius by $r_c$, the modified geodesic length is given by~\cite{Griguolo:2025kpi}.
\begin{align}
\begin{split}
    \ell_{T\bar T}(\Delta \tau; r_c, r_h)&
=2\,\text{sinh}^{-1}\!\left(
\frac{\sqrt{r_c^2 - r_h^2}}{r_h}
\sin\!\left(\frac{r_h \Delta \tau}{2}\right)
\right)
\approx 2\, \text{sinh}^{-1}\bigg(\alpha \,\text{exp}{\Big(\frac{\ell_{\textrm{ren}}^{\text{JT}}}{2}}\Big)\bigg)\,
\label{6.48u},
\end{split}
\end{align}

\noindent
where,
\begin{align}
    \ell_{\mathrm{ren}}^{\text{JT}}(\Delta \tau)
\propto
2 \log \sin\!\left(\frac{r_h \Delta \tau}{2}\right)\,,\quad \alpha=\frac{r_c}{r_h},\,\,r_h=\frac{2\pi}{\beta}.
\end{align}

\noindent
Therefore, the matrix elements of the deformed theory can be obtained by substituting \eqref{6.48u} in the exponential and using (\ref{new3})  as follows \footnote{we will drop suffixes and call $\ell_{\text{ren}}^{\text{JT}}\to\ell$. },
\begin{align}
\begin{split}
  \int_{-\infty}^{\infty}d\ell\, \psi_{E}^{\lambda}(\ell)\psi_{E'}^{\lambda}(\ell)\,e^{-\Delta \ell_{T\bar T}}&=\Bigg(\frac{r_c}{r_h}\Bigg)^{-2\Delta}\frac{\left|\Gamma\left(\Delta+
  (i\sqrt{s_1^2(1-2\lambda s_1^2)}\pm i\sqrt{s_2^2(1-2\lambda s_2^2)})\right)\right|^2}{2^{2\Delta+1}\Gamma(2\Delta)}\\
  &=\mathcal{M}_{\Delta}^\lambda(E,E')\,.\label{6.50y}
\end{split} 
\end{align}
    Here we have used the notation $\Gamma(\Delta+(a\pm b)) = \Gamma(\Delta+(a+ b))\Gamma(\Delta+(a- b))$. Now, we proceed to calculate the expectation or one-point function of the length $\langle\ell(t)\rangle$.

\section{Expectation of ERB length}\label{sec6}

Following the arguments in \cite{Iliesiu:2021ari}, the length of the Einstein-Rosen Bridge (ERB) is defined as, 
\begin{equation}
    \langle \ell \rangle = \lim_{\Delta\rightarrow 0}\left\langle\sum_{\gamma}\ell_\gamma e^{-\Delta\ell_\gamma}\right\rangle
\end{equation}
where $\gamma$ labels the non-self-intersecting geodesics, $\langle\cdots\rangle$ represents summing over surfaces of arbitrary topologies while evaluating the gravitational path integral, and $\Delta$ acts as a regulator. This definition of length can be related to the two-point function $\langle\chi\chi\rangle$ of the operator of conformal dimension $\Delta$ inserted on each side of the two-sided black hole. We do not have to worry about divergences in this quantity since the time-dependent piece that we are interested in is finite \cite{Iliesiu:2021ari}. Hence, we study the quantity $\langle \ell(t) \rangle - \langle \ell(0) \rangle$, which is independent of the regularization procedure. $\langle \ell(t) \rangle$ can be computed by taking the $\Delta$ derivative of the two-sided correlation function as given below:
\begin{equation}
    \langle\ell(t)\rangle = - \lim_{\Delta \rightarrow 0} \frac{\partial }{\partial\Delta}\big\langle \chi_L(t)\chi_R(0)\big\rangle_{\text{non-int}}\,.
\end{equation}
In { JT gravity}, the integral over all metrics reduces to an integral over the boundary wiggles with measure $\mathcal{D}(\mathcal{W})$. The two-point function then can be cast as \cite{Iliesiu:2021ari},
 \begin{align}
     \textrm{Tr}_{\beta}\langle\chi(x_1)\chi(x_2)\rangle_{\textrm{non-int}}=\sum_{g}e^{S_0(1-2g)}\int_{\frac{\mathcal{T}_{g,1}}{Mod(\mathcal{M}_{g,1})}}\omega\int \mathcal{D}(\mathcal{W})\,e^{-I_{JT},bdy(\mathcal{W})}\sum_{\gamma}e^{-\Delta l_{\gamma}}
 \end{align}
where $$\omega=\sum_{j=1}^{3g-3+n}db\wedge d\tau$$ is the usual Weil-Peterson symplectic form defined on the moduli space of hyperbolic Riemann surfaces. Using the same analogy for $T\bar{T}$ deformed theory, the trumpet wavefunction in terms of $\psi_{E,\lambda}(\ell)$, with a geodesic of length $\ell$ can be written as,
\begin{align}
    \psi^{\textrm{trumpet},x}_{\lambda}(\ell,b)=\begin{minipage}[h]{0.15\linewidth}
	\vspace{-0.1 cm}
	\scalebox{1.3}{\includegraphics[width=\linewidth]{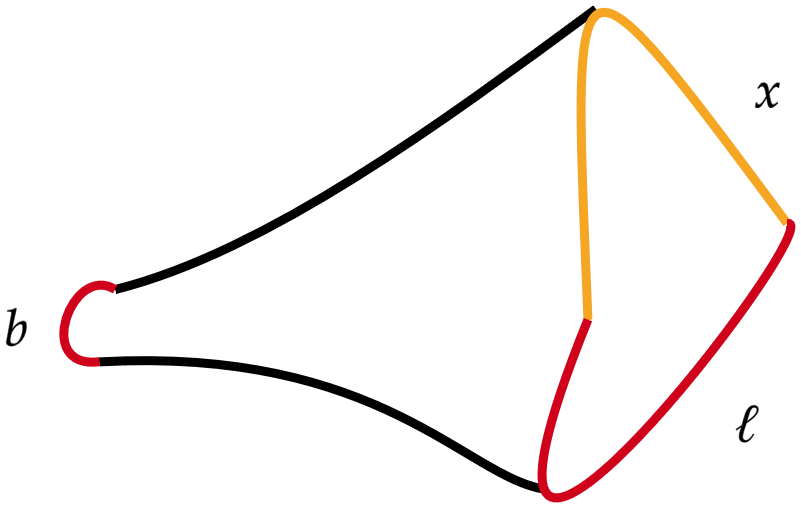}}
\end{minipage}\hspace{0.8 cm}=\int_0^\infty dE \rho^{\textrm{trumpet}}_{\lambda}(b,E)\psi_{E,\lambda}(\ell)\,e^{-xE},
\end{align}
where,  \begin{align}\rho_\lambda^{\text{trumpet}}(b,E) = (1-4\lambda E)\frac{\cos\left(b\sqrt{E(1-2\lambda E)}\right)}{2\pi\sqrt{E(1-2\lambda E)}}\end{align} is the deformed trumpet density of states. Now, the genus-$g$ contribution can be written compactly as,
\begin{align}
\begin{split}
     \textrm{Tr}_{\beta}(\chi(x_1)\chi(x_2))\rangle_{\textrm{non-int}}&\sim e^{S_0(1-2g)}\int_{-\infty}^{\infty} d\ell\, e^{\ell}\int_{0}^\infty db_1 b_1db_2 b_2\psi^{\textrm{trumpet},x}_{\lambda}(\ell,b_1)\psi^{\textrm{trumpet},\beta-x}_{\lambda}(\ell,b_2)\,e^{-\Delta \ell}\\&\times\Bigg[\textrm{Vol}_{g-1,2}(b_1,b_2)+\sum_{h\geq0}\textrm{Vol}_{g-h,1}(b_1)\textrm{Vol}_{h,1}(b_2)\Bigg]\,.
     \end{split}
\end{align}
\noindent
Therefore, the two-point function for the $T\bar{T}$ deformed theory can be written in terms of the two-point density correlator and the squared matrix element of operator insertions as,
\begin{align}
\label{two_point_fn}
    \begin{split}
        \big\langle \text{Tr}_\beta(\chi(x_1)\chi(x_2))\big\rangle_{\text{non-int.}} = \frac{4e^{-S_0}}{Z_\lambda^{\text{disk}}(\beta)} \int_0^\infty dE_1 dE_2 \big\langle \rho(E_1) \rho(E_2) \big\rangle_{\lambda} e^{-E_1(x_1-x_2)-E_2(\beta-x_1+x_2)}\mathcal{M}^\lambda_\Delta(E_1,E_2)
    \end{split}
\end{align}
where $\mathcal{M}^\lambda_\Delta(E_1,E_2)$ takes the form given in equation \eqref{6.50y} with $s_{1,2} \equiv \sqrt{E_{1,2}}$. Since there is a unique geodesic for a disk, we see if $\langle\chi(x_1)\chi(x_2) \rangle_{\text{non-int.}}^{\Delta\rightarrow 0} = 1$ at the leading order as a consistency check. We have,
\begin{align}
    \begin{split}
        \mathcal{M}_0^\lambda(s_1,s_2) &= \frac{\delta\left(s_1-s_2\right)}{4(1-4\lambda s_1^2)\,r\left(\sqrt{s_1^2(1-2\lambda s_1^2)}+\sqrt{s_2^2(1-2\lambda s_2^2)}\right)},
        \,\, \hspace{1 cm}r(s) = \frac{s}{2\pi^2}\sinh(2\pi s)\,.\label{6.8m}
    \end{split}
\end{align}
Here, we have used the following properties of gamma functions
\begin{equation}
    |\Gamma(bi)|^2 = \frac{\pi}{b \sinh(\pi b)}\hspace{0.5cm};\hspace{0.5cm}{\bf B(z,-z)} = \lim_{\epsilon\rightarrow 0}\frac{\Gamma(z)\Gamma(-z)}{\Gamma(\epsilon)} = 2\pi\delta(z)
\end{equation}
where ${\bf B(x,y)}$ is the beta function. Hence we have,
\begin{align}
    \begin{split}
        \big\langle\chi(x_1)\chi(x_2) \big\rangle_{\text{non-int.}}^{\Delta\rightarrow 0}\big|_{\text{LO}} &= \frac{4e^{-S_0}}{Z_\lambda^{\text{disk}}(\beta)} \int_0^\infty  ds_1 \,ds_2\,4s_1 s_2 \rho_\lambda^{\text{disk}}(s_1^2) \rho_\lambda^{\text{disk}}(s_2^2)e^{-(s_1^2(x_1-x_2)+s_2^2(\beta-x_1+x_2))}\mathcal{M}^\lambda_0(s_1,s_2)\,,\\
        &= \frac{1}{Z_\lambda^{\text{disk}}(\beta)} \int_0^\infty ds_1 2 s_1 \rho_\lambda^{\text{disk}}(s_1^2) e^{-\beta s_1^2} =1\,.
    \end{split}
\end{align}
We perform this check to validate the matrix element we found in \eqref{6.50y}. Now, we discuss the growth rate of the Einstein-Rosen bridge (ERB). This can be calculated from the one-point function of $\ell(t)$ in the deformed theory. We also study the saturation of the interior growth in $T\bar{T}$ deformed setup by using the non-perturbative kernel and $T\bar{T}$ deformed matrix element.

\subsection*{Time-dependence of ERB length}
We study the time dependence of the length of the Einstein-Rosen bridge using the following relation,
\begin{equation}
    \big\langle \ell(t) \big\rangle \equiv \big\langle \text{Tr}_\beta (\hat{\ell}) \big\rangle \equiv -\lim_{\Delta \rightarrow 0} \frac{\partial \left\langle \text{Tr}_\beta\left(\chi(\frac{\beta}{2}+it)\chi(0)\right) \right\rangle_{\text{non-int}} }{\partial \Delta}\,.
\end{equation}
We define, \begin{align}
    \omega \equiv s_1-s_2\,\, \text{  and   } \,\,\,\,\,\bar{s} = \frac{s_1+s_2}{2}
    \label{7.12y}\,.
\end{align}
From  \eqref{two_point_fn} we can see that the only $\Delta$ dependent term in the two-point function is $\mathcal{M}_\Delta^\lambda$, hence we take the derivative inside the integral:
\begin{align}
    \begin{split}
        \textcolor{black}{-\lim_{\Delta\rightarrow 0} \frac{\partial \mathcal{M}_\Delta^\lambda}{\partial \Delta} =\#\delta(\omega) - \frac{1}{16\pi^2 \,r(\bar{s})\,r(\frac{\omega}{2})}\,+2 
        \log \left(\frac{r_c}{r_h}\right)\mathcal{M}^{\lambda}_0(s_1,s_2)\,.}\label{7.13e}
    \end{split} 
\end{align}
Since $\#\delta(\omega)$ only gives a time-independent contribution, the expression for length becomes,
\begin{align}
    \begin{split}
        \big\langle \ell(t) \big\rangle_{\lambda} = &\text{\,const} - \frac{e^{-S_0}}{4\pi^2  Z_\lambda^{\text{disk}}(\beta)} \int_0^{\infty} ds_1 ds_2\,s_1s_2\Bigg[\frac{\big\langle \rho(s_1) \rho(s_2) \big\rangle_{\lambda}}{r(\bar{s})r(\frac{\omega}{2})}+2\log\left(\frac{r_c}{r_h}\right)\textcolor{black}{{\big\langle \rho(s_1) \rho(s_2) \big\rangle_{\lambda}}\mathcal{M}^{\lambda}_0(s_1,s_2)}\Bigg]  \,\,\\& \hspace{7.2 cm}\times\text{exp}\left[-\beta\left({\bar{s}^2} + \frac{\omega^2}{4}\right)-i\bar{s}\omega t\right]\,,
        \label{5.8ke}
    \end{split}
\end{align}
where $\big\langle \rho(s_1) \rho(s_2) \big\rangle_{\lambda}$ is the $T\bar{T}$ deformed correlator and takes the form as given in \eqref{5.23c}. The contact term gives a time-independent contribution to the integral and hence can be omitted. We break the above expression into two parts, 
\begin{align}
    \begin{split}
     &   I^{(1)}= \text{\,const} - \frac{e^{-S_0}}{4\pi^2  Z_\lambda^{\text{disk}}(\beta)} \int_0^{\infty} ds_1 ds_2\,s_1s_2\,\,\frac{\big\langle \rho(s_1) \rho(s_2) \big\rangle_{\lambda}}{r(\bar{s})r(\frac{\omega}{2})}\text{exp}\left[-\beta\left({\bar{s}^2} + \frac{\omega^2}{4}\right)-i\bar{s}\omega t\right]\,,\\&
     I^{(2)}=  - \frac{e^{-S_0}}{4\pi^2  Z_\lambda^{\text{disk}}(\beta)} \int_0^{\infty} ds_1 ds_2\,s_1s_2\,\,\Big(2\log\left(\frac{r_c}{r_h}\right)\textcolor{black}{{\big\langle \rho(s_1) \rho(s_2) \big\rangle_{\lambda}}\mathcal{M}^{\lambda}_0(s_1,s_2)}\Big)\text{exp}\left[-\beta\left({\bar{s}^2} + \frac{\omega^2}{4}\right)-i\bar{s}\omega t\right]\,. \label{new4}
    \end{split}
\end{align}
\noindent
We start by evaluating $I^{(1)}\,.$ We will evaluate it the limit $|s_1-s_2|\ll1\,.$
\begin{figure}
\centering\scalebox{0.42}{\includegraphics{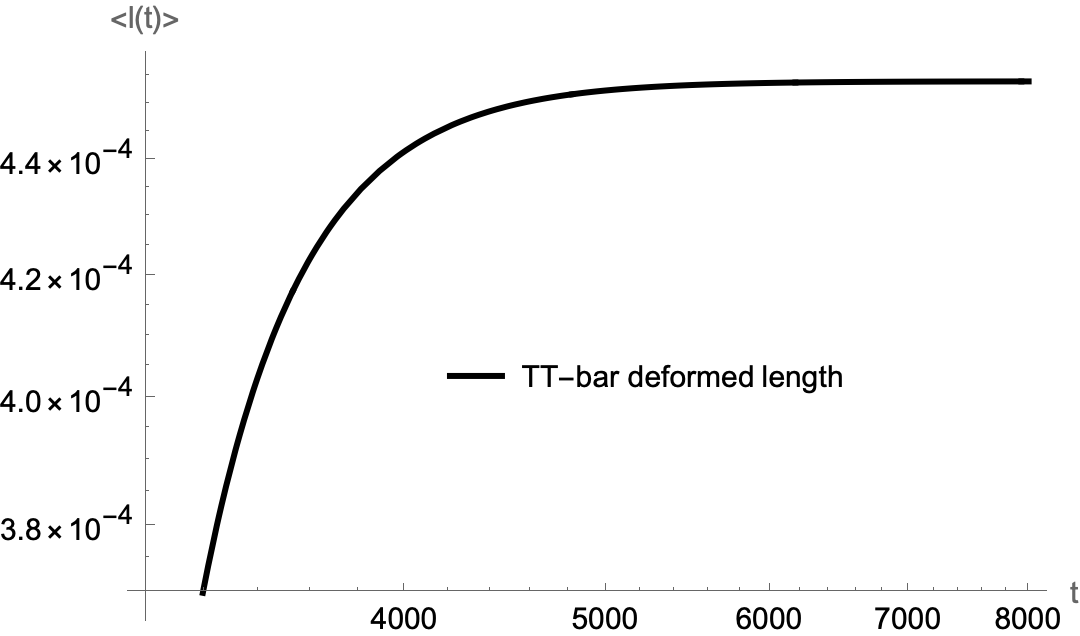}}
\caption{ The growth and saturation of ERB length for $T\bar{T}$ deformed theory.}
\label{fig_3}
\end{figure}
\newpage
\noindent
\textbf{The $\omega$ integral:} 
The $\omega$ integral is a part of variable change from $\{s_1,s_2\}$ to $\{s,\omega\}$. \textcolor{black}{We are mainly interested in large time regime ($t>t_{*}$). Therefore, $\exp[-i\bar{s}\omega t]$ will be dominated by $\omega\sim 1/t$} and we can implement this by scaling $\omega=\frac{\hat{\omega}}{t}$. Now, the $\omega$ integral coming from $I^{(1)}$ becomes\footnote{We drop the bar from $s$ and call it only $s$.},
\begin{align}
    \begin{split}
\mathscr{I}(s)=\mathcal{N} t\int_{-\infty}^{\infty}d\hat{\omega}\,e^{-i s\hat{\omega}}\frac{x}{\hat{\omega}^2 }\Bigg[\underbrace{{1}}_{\text{semiclassical part}}-\underbrace{\textcolor{black}{{\frac{t^2}{8\pi^2\hat\rho^2\hat\omega^2 s^2}+\frac{ t^2e^{-2\lambda\omega^2}}{8\pi^2\hat\omega^2 s^2\hat\rho^2}\textrm{cos}\Bigg(\frac{4\pi\hat{\rho}\hat{\omega} s}{t}\Bigg)}}}_{\text{non-perturbative part}}\Bigg]\label{6.9a}
    \end{split}
\end{align}
where, 
\begin{align}x=-8 \pi ^3 \text{csch}(2 \pi  s) \left(2 \pi  \lambda  s^3 \coth (2 \pi  s)+7 \lambda  s^2-1\right)\,.\end{align}
In \eqref{6.9a}, the first term in the parentheses comes from the semiclassical disconnected part, and the rest of the terms come from the non-perturbative contribution \footnote{\noindent
\textcolor{black}{ One can notice that the expectation value of length of the baby universe as also advocated in ~\cite{Sato:2025uli}, remains unchanged in the deformed theory. This can be understood as follows.
In the computation of $\langle \ell(t) \rangle$, the term proportional to $e^{-S_0}$ provides the leading contribution associated with geometries in which the Lorentzian evolution emits a single baby universe. In the context of JT gravity, it was shown in~\cite{Sato:2025uli} that the expectation value of the length of the baby universe is directly proportional to the coefficient of this $e^{-S_0}$ term. In particular, both this expectation value and the corresponding contribution to $\langle \ell(t) \rangle$ scale as $T^3$, where $T$ denotes the duration of Lorentzian evolution. Upon turning on the $T\bar{T}$ deformation, we observe that the $e^{-S_0}$ term in \eqref{6.9a} does not receive any corrections depending on the deformation parameter $\lambda\,.$ Since the expectation value of the length of the baby universe is controlled precisely by this term, it follows that it remains unaffected by the deformation, and therefore coincides with that of undeformed JT gravity.}}.
After performing the Fourier transform, and using the notation $\rho_{\lambda}(s)=e^{S_0}\hat{\rho}_{\lambda}(s) \text{ and } \,\, t=e^{S_0}\hat{t}$, we get,
\begin{align}
    \begin{split}
\mathscr{I}(s)=&384 \pi ^5 \sqrt{2} \lambda  \hat\rho_{\lambda}(s) ^4 s^7 (2 \pi  s \coth (2 \pi  s)+7)-384 \pi ^5 \sqrt{2}\hat\rho_{\lambda}(s) ^4 s^5-128 \pi ^{3/2} \sqrt{\lambda } \hat\rho_{\lambda}(s)  s^2 t^2 e^{-\frac{\left(64 \pi ^2 \hat\rho_{\lambda}(s)^2+1\right) s^2}{8 \lambda }} \\&\sinh \left(2 \pi  \hat\rho_{\lambda}(s)  s^2/\lambda \right)+8 \sqrt{\pi } \sqrt{\lambda } \left(64 \pi ^2 \hat\rho_{\lambda}(s)^2+1\right) s^2 t^2 e^{-\frac{\left(64 \pi ^2 \hat\rho_{\lambda}(s)^2+1\right) s^2}{8 \lambda }} \cosh \left(2 \pi  \hat\rho_{\lambda}(s)  s^2/\lambda \right)\\&-4 \sqrt{\pi } \sqrt{\lambda } (8 \pi  \hat\rho_{\lambda}(s) +1)^2 s^2 t^2 e^{-\frac{(8 \pi  \hat\rho_{\lambda}(s) s+s)^2}{8 \lambda }}-4 \sqrt{\pi } \sqrt{\lambda } (1-8 \pi\hat\rho_{\lambda}(s) )^2 s^2 t^2 e^{-\frac{(1-8 \pi  \hat\rho_{\lambda}(s) )^2 s^2}{8 \lambda }}\\&-\sqrt{2} \pi  s t^2 \Big((1-8 \pi  \hat\rho_{\lambda}(s) )^3 \left(-s^2\right)+2 s^2-{(8 \pi \hat\rho_{\lambda}(s) +1)} s^2(8 \pi  \hat\rho_{\lambda}(s)  +1)^2\Big)\\& +\sqrt{2} \lambda  t^2 \Big(\pi  s^3 \Big[(1-8 \pi  \hat\rho_{\lambda}(s))^3 \left(-s^2\right)+2 s^2\\&-{(8 \pi  \hat\rho_{\lambda}(s) +1)} s^2(8 \pi\rho_{\lambda}(s) +1)^2(2 \pi  s \coth (2 \pi  s)+7)\Big]+12 \pi  \left(16\pi \hat\rho_{\lambda}(s)\right) s\Big)\,.
 \end{split}
\end{align}
Since this function is very complicated, one can only perform the final integral numerically.
However, if we want to compute the ERB length at a large time, then analytical computation is possible. The remaining $s$ integral then can be performed, and we get,
\begin{align}
    \begin{split}
        I^{(1)} &= \text{const} - \frac{\hat{t} e^{S_0}}{4\pi^2  Z_\lambda^{\text{disk}}(\beta)} \int_0^\infty d{s}\, {s}^2\,{\hat \rho_\lambda(s)^2 } 
        \times\,\mathscr{I}({s})e^{-\beta s^2}\,,\, \textrm{erf}(t^2)\overset{t\sim \infty}{\sim} 1-\frac{e^{-t^4}}{\sqrt{\pi } t^2}\,\,,\hat{t}=te^{-S_0}\label{5.8k}
    \end{split}
\end{align}
\noindent
\textcolor{black}{Here, the lower limit of the $s$-integral, $s_*(t)$, is fixed from the argument inside the cosine term in \eqref{6.9a} for the convergence of the Fourier transform, which we explain below. 
We close the contour in the lower half plane, such that the arc at infinity does not contribute \footnote{For pure JT gravity, there was a suppression factor $\sim e^{-\beta s^2}$, due to which the cancellation of the contribution from the poles at $\hat\omega=0$ between the semiclassical piece and non-perturbative contribution was non-trivial. But in $T\bar{T}$ deformed case, we do not have that sort of suppression factor, as it was already canceled. So, the closure of the contour should be in the lower half-plane, and that fixes $s_*(t)$. We always close the contour in such a way that the arc at complex infinity does not contribute. }.
Now, one should note that if $\lambda=0$, the non-perturbative part and the semiclassical part cancel. For $\lambda\neq 0$, something interesting happens. Expanding the exponential in the non-perturbative part in \eqref{6.9a}, we get,
\begin{align}\exp(-2\lambda\omega^2)\approx1-2\lambda\omega^2+\mathcal{O}(\omega^4).
\end{align} \textit{Now, all the higher order terms except the term proportional to $\omega^2$ are regular, as the pole at $\omega=0$ cancels.} Therefore, we only focus on the lower half plane with non-vanishing contribution for $(1+4\lambda s^2)\hat{t}>\pi \hat{\rho}_{\lambda}(s)$, which in turn says the lower limit of $s$-integral should be fixed by, 
\begin{align} 
\hat t=\pi\,\frac{\hat\rho_{\lambda}(s_*(\hat t))}{1+4\lambda s_*(\hat t)^2}. 
\end{align}
Here on we set $\lambda=-|\lambda|$. Now we try to solve for $s_*(t)$ and get the following},\\
\begin{align}s_*(\hat{t})=\frac{1}{2\sqrt{\pi\lambda}} \left(\pi-\gamma_4\right)^{\frac{1}{2}}\,,\quad \gamma_4=\sqrt{\pi ^2-2 \lambda  \sinh ^{-1}\left(\frac{\hat{t}}{2 \pi }\right)^2}\,.\end{align}\\So the leading order term at large $t$ is given by, 
\begin{align}
    \begin{split}
        I^{(1)} &= \text{const} + \frac{\hat{t}^3 e^{3S_0}}{4\pi^2  Z_\lambda^{\text{disk}}(\beta) } \int_{s_{*}(\hat{t})}^\infty ds\,{\hat \rho_\lambda(s)^2 }
        s^2\\&\hspace{3 cm}\times\Bigg(-\frac{4 \sqrt{2} \pi ^{3/2} e^{\beta  \left(-s^2\right)} \text{csch}(2 \pi s) \left(2 \pi  |\lambda|  s^3 \coth (2 \pi  s)+7 |\lambda|  s^2-1\right)}{s^2\hat{\rho}_{\lambda}^2(s)}\Bigg)\,.\label{5.9 m}
    \end{split}
\end{align}

\begin{figure}
\scalebox{0.35}{\includegraphics{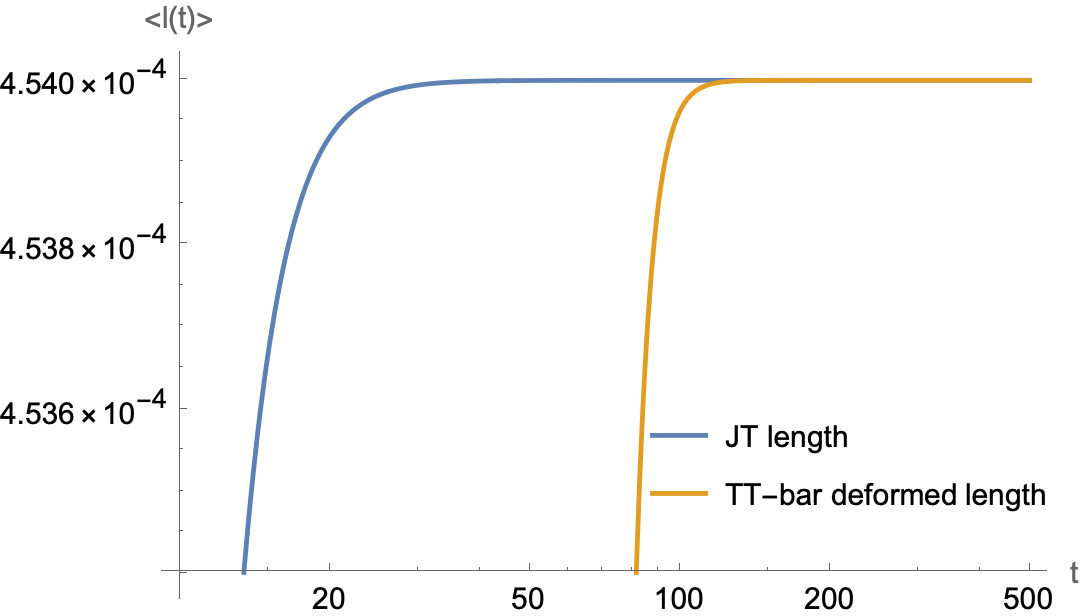}}\hspace{1 cm}
\scalebox{0.22}{\includegraphics{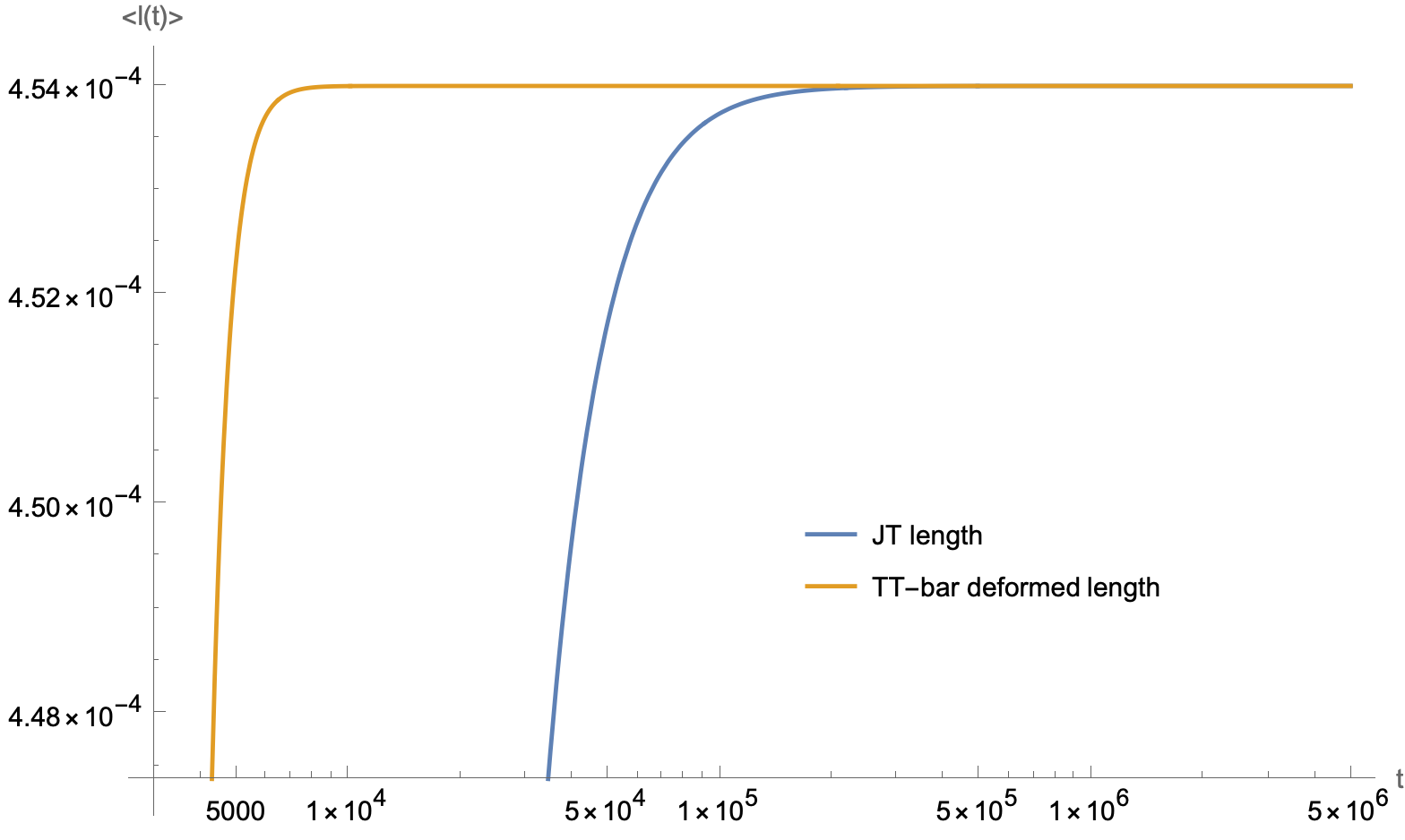}}
\caption{Plots showing the growth of the ERB for JT gravity and $T\bar T$ deformed JT gravity. Both graphs are plotted in the Log-Log scale. The first graph shows that JT length saturates later than the $T\bar T$ deformed case, while the latter shows the opposite case, where the JT length saturates faster, indicating a transition in between depending on the value of $\beta\,.$ The first plot (LHS) is for $\beta=15$, and the latter (RHS) is for $\beta=60$. Both of them are plotted for $|\lambda|=0.1$. Also note that, $r_h=\frac{2\pi}{\beta}$ and $r_c^2=\frac{2\pi}{|\lambda|}$ and we have set $G_N=1\,.$}\label{fig_4}
\end{figure}

\noindent
\textbf{\textit{Contribution from the logarithmic term}}
\\
Next we evaluate the $I^{(2)}$ as defined in (\ref{new4}). We proceed in a similar way as discussed earlier for $I^{(1)}\,.$ The integral contribution to the length from the log-term in \eqref{7.13e},\\
\begin{align}
\begin{split}
I^{(2)}
&= - \frac{e^{-S_0}}{4\pi^2  Z_\lambda^{\text{disk}}(\beta)} \frac{\log\!\left(\frac{r_c}{r_h}\right)}{\sqrt{2}}
\int_{s_{*}(t)}^{\infty} 
ds\,e^{-\beta s^2}\rho_{\lambda}(s)^2 s^2
\frac{\pi^2}{s \sinh(2\sqrt{2}\pi s)}\,,
\\&\sim  - \frac{e^{-S_0}}{4\pi^2  Z_\lambda^{\text{disk}}(\beta)} \Bigg[ \frac{\log\!\left(\frac{r_c}{r_h}\right)}{32 \sqrt{2}\,\pi^2\,\beta^{3/2}}
\Bigg[
 \sqrt{\beta}\,
\exp\!\Big(
-\frac{1}{4\lambda^2}
\Big[
\frac{\beta(1-\gamma_4)}{\pi}
- 4\,\gamma_5\sqrt{\pi}\,|\lambda|\,
\sqrt{\pi - \gamma_4}
\Big]
\Big)\\&\hspace{5.5cm} -\gamma_5\,\pi^{3/2}
\, e^{\frac{(6 - 4\sqrt{2})\pi^2}{\beta}} \Big\{1-
\erf\!\Big(
\frac{1}{\sqrt{\beta}}
\Big[
\frac{\beta}{2|\lambda|}
\sqrt{1 - \frac{\gamma_4}{\pi}}
+ \pi\,\gamma_5
\Big]
\Big)\Big\}\Bigg]\,.\label{7.21r}
\end{split}
\end{align}
\noindent
Here, $\gamma_5=2-\sqrt{2}\,.$
Then after performing the integral asymptotically, i.e., replacing the $\textrm{sinh}$ functions by exponential and adding \eqref{5.9 m} and \eqref{7.21r}, we get,
\hfsetfillcolor{gray!10}
\hfsetbordercolor{white}
\begin{align}\begin{split}
\tikzmarkin[disable rounded corners=true]{kol}(2.6,-1.08)(-0.8,1.2)
   \hspace{-0.7 cm} \big\langle \ell(t) \big\rangle_{\mathcal{O}({\lambda})}\approx&\, \text{Const.} -e^{S_0}\hat{t}^2\frac{\exp \left(-\frac{\pi ^2 \left(\beta ^2+2 \pi ^2 |\lambda| \right)}{\beta ^3}-\frac{\log ^2(\hat{t}) \left(2 \pi ^2 \beta +|\lambda| \log (\hat{t}) \left(\beta  \log (\hat{t})+2 \pi ^2\right)\right)}{8 \pi ^4}\right)}{\sqrt{2} \pi ^{7/2} \beta ^2 }\\&\times\Bigg[\Big(-e^{\frac{\left(\frac{\beta  \log (\hat{t})}{2 \pi }+\pi \right)^2}{\beta }} \text{erfc}\left(\frac{\beta  \log (\hat{t})+2 \pi ^2}{2 \pi  \sqrt{\beta }}\right)\Big(8 \pi ^4 \Big(\beta ^3-11 \beta ^2 |\lambda| -14 \pi ^2 \beta  |\lambda| +2 \pi ^4 |\lambda|)\Big)\\&+\beta ^3|\lambda|  \log ^3(\hat{t}) \left(\beta  \log (\hat{t})+2 \pi ^2\right)\Big)+2 \sqrt{\pi } \sqrt{\beta }|\lambda|\Big(-20 \pi ^4 \beta \\&+\beta  \log (\hat{t}) (14 \pi ^2 \beta +\beta  \log (\hat{t}) \left(\beta  \log (\hat{t})+2 \pi ^2\right)-4 \pi ^4)+8 \pi ^6\Big)+{I}^{(2)}(\hat{t})\Bigg]\,.
        \tikzmarkend{kol}
\end{split}
\end{align}
\noindent
\textcolor{black}{We observe that the growth of the length in the $T\bar{T}$-deformed theory also exhibits saturation  as illustrated in Fig.~\eqref{fig_3}}. A direct comparison between the $T\bar{T}$-deformed case and that of the JT gravity is shown in Fig.~\eqref{fig_4}, where the distinction becomes apparent throughout the evolution.

\textcolor{black}{An intriguing feature emerges when varying the inverse temperature parameter $\beta$. For relatively smaller values, such as $\beta = 15$, the JT gravity result saturates more rapidly than its $T\bar{T}$-deformed counterpart. \textit{In contrast, for larger values like $\beta = 60$, this behavior is reversed, with the $T\bar{T}$-deformed theory exhibiting earlier saturation. This qualitative change strongly suggests the presence of a nontrivial crossover behavior.}
Such a transition bears resemblance to a lower-dimensional analog of a Hawking-Page type phase transition, at least within a perturbative regime characterized by the good  sign of the deformation parameter $\lambda$. This indicates that the $T\bar{T}$ deformation can significantly modify the late-time dynamics, potentially altering the effective thermodynamic phases of the system}. A more systematic study this crossover (which will also depend on the magnitude of $\lambda$) is left for future work.\\

\noindent
\textbf{A possible connection with Krylov complexity}
\noindent
\textcolor{black}{As the length of ERB saturates for the $T\bar{T}$ deformed theory, similar to JT gravity, it is natural to expect some connection with \textit{Krylov complexity} following \cite{Kar:2021nbm, Balasubramanian:2024lqk} which initially grows and then saturates.} The  Krylov complexity of an operator $\mathcal{O}(t)$ is given by \cite{PhysRevX.9.041017},
\begin{align}
    C_{E}(t)=\sum_{n=0}^{D(E)}n|\phi_{E,n}(t)|^2\,.\label{6.22c}
\end{align}
$D(E)$ is the dimension of the Krylov space. Now the $\phi_{E,n}(t)s$ can be found by solving the following discrete Schrodinger equation in the Krylov basis,
\begin{align}
 \dot{\phi}_{E,n}(t) = b^{E}_{n+1}\phi_{E,n+1}(t)-b^E_n\phi_{E,n-1}(t)
\end{align}
where, $b^{E}_n$ are the Lanczos coefficients \cite{viswanath1994recursion}. 
This exact equation is valid in any quantum state. Hence, we proceed to compute the \textit{K-complexity} initial growth using the moment method as follows.
 The deformed matrix element as obtained in \eqref{6.50y} for our case  with $s_1=\sqrt{E_1},\,s_2=\sqrt{E_2}$ is given by,
\begin{align}
    \begin{split}\mathcal{M}_{\Delta}^\lambda(E,E')&={\Bigg(\frac{r_c}{r_h}\Bigg)^{-2\Delta}}\frac{\left|\Gamma\left(\Delta+
  (i\sqrt{E_1(1-2\lambda E_1)}\pm i\sqrt{E_2(1-2\lambda E_2)})\right)\right|^2}{2^{2\Delta+1}\Gamma(2\Delta)}\,.
    \end{split}
\end{align}
Now, at the leading order of the two-point correlator (the disconnected piece is enough for showing the growth), $\langle\rho\rho\rangle$ in the semiclassical limit is given by,
\begin{align}
    \begin{split}
        \rho_{\textrm{def}}(E,\omega)=\rho_{\lambda}^{\text{disk}}\left(E+\frac{\omega}{2}\right)\rho_{\lambda}^{\text{disk}}\left(E-\frac{\omega}{2}\right),\,\quad \omega=E_1-E_2, \,E=\frac{E_1+E_2}{2}\,.
    \end{split}
\end{align}
Explicitly, it takes the following form,
\begin{align}
    \begin{split}
        \rho_{\textrm{def}}(E,\omega)=\frac{1}{4\pi^2}\left(1-4\lambda\left(E+\frac{\omega}{2}\right)\right)\sinh\left[2\pi\sqrt{\left(E+\frac{\omega}{2}\right)\left(1-2\lambda\left(E+\frac{\omega}{2}\right)\right)}\right]\times (\omega\to -\omega)\,.
    \end{split}
\end{align}
Focusing on the small moments, one can approximate the two-point correlator as,
\begin{align}
    \begin{split}
        \rho_{\textrm{def}}(E,\omega) \to (1-8\lambda E)\exp\left(4\pi \sqrt{{E} (1-2 {E} \lambda )}\right)
    \end{split}
\end{align}
and,
\begin{align}
    \begin{split}
        \mathcal{M}_{\Delta}^\lambda(E,\omega)&\xrightarrow[]{} \Bigg(\frac{r_c}{r_h}\Bigg)^{-2\Delta}\\&\times\frac{\left|\Gamma\left(\Delta+i\left(2\sqrt{E(1-2E\lambda)}+\mathcal{O}(\frac{\omega^2}{E^2})\right)\right)\Gamma\left(\Delta+i\left(\frac{(1-4\lambda E)}{\sqrt{1-2E\lambda}}\right)\frac{\omega}{2\sqrt{E}}+\mathcal{O}(\omega^2/E^{3/2})\right)\right|^2}{2^{2\Delta+1}\Gamma(2\Delta)}.
    \end{split}
\end{align}
Now, using the following identity \cite{Kar:2021nbm},
\begin{align}
    |\Gamma(\Delta+ir)|^2=\frac{\pi r}{\sinh(\pi r)}\prod_{k=1}^{\Delta-1}(r^2+k^2),
\end{align}
we can write the matrix element as (for a qualitative analysis, it is not a real loss to restrict the integer $\Delta$ and),
\begin{align}
    \begin{split}
        \mathcal{M}_{\Delta}^\lambda(E,\omega)&\xrightarrow[]{} \textcolor{black}{\Bigg(\frac{r_c}{r_h}\Bigg)^{-2\Delta}}\frac{1}{\sinh\left(2\pi\sqrt{E(1-2E\lambda)}\right)\sinh\left(\pi \frac{(1-4\lambda E)}{\sqrt{1-2E\lambda}}\frac{\omega}{2\sqrt{E}}\right)}\,,\\ &
        \sim \Bigg(\frac{r_c}{r_h}\Bigg)^{-2\Delta}\exp\left(S(E)-\frac{\beta_{E}\omega}{2}\right),
    \end{split}
\end{align}
where,
\begin{align}
    S(E)=2\pi \sqrt{E(1-2E\lambda)},\,\,\beta_{E}=\pi \frac{1-4\lambda E}{\sqrt{E(1-2E\lambda)}}\equiv \frac{\partial S(E)}{\partial E}\,.\label{134k}
\end{align}
In \eqref{134k}, $S(E)$ is the microcanonical entropy of finite cutoff JT gravity above extremality. Now, the average energy moments are given by \cite{Kar:2021nbm},
\begin{align}
    \begin{split}
        m_{2n}^{E}&\xrightarrow[]{}\frac{1}{\mathcal{N}(E)} \int _{-2E}^{2E} d\omega\,\omega^{2n} \rho_{\textrm{def}}(E,\omega)\,\mathcal{M}_{\Delta}^\lambda(E,\omega)\,,\\&
        =\Bigg(\frac{r_c}{r_h}\Bigg)^{-2\Delta}\frac{1}{\int_{-2E}^{2E}d\omega\,\exp\left(-\beta_{E}\omega/2\right)}\int_{-2E}^{2E}d\omega\, e^{2 n\log \omega} \exp\left(-\frac{\beta_E \omega }{2}\right)\,.
    \end{split}
\end{align}
One can do the integral using the saddle point approximation, which is valid in the parametric region $\beta_{E}^{-1}\ll \omega \ll E$. The saddle point condition reads,
\begin{align}
    \frac{2n}{\omega_{*}}=\frac{\beta_{E}}{2}\implies \omega_{*}=\frac{4n}{\beta_{E}}\,.
\end{align}
Therefore, the moments become,
\begin{align}
    \begin{split}
        m_{2n}^{E}=\frac{1}{\mathcal{N}(E)}\left(\frac{4n}{\beta_E}\right)^{2n}e^{-2n+\cdots}\to e^{n\left(\log (4n/\beta_{E})^2-2\right)}\,.
    \end{split}
\end{align}
For large $n$ we can approximate it to,
\begin{align}
    m_{2n}^E\sim  e^{n[\log (4n/\beta_{E})]^2}\,.
\end{align}
Therefore, the Lanczos coefficient is given by,
\begin{align}
    b_{n}^{E}\sim \frac{n}{\beta_{E}},\,\,\,\,\beta_{E}=\pi \frac{1-4\lambda E}{\sqrt{E(1-2E\lambda)}}\,.
\end{align}
Following \cite{Kar:2021nbm},  using the definition of Krylov complexity \eqref{6.22c} we have,
\begin{align}
    C_{E}(t)\sim e^{\frac{2\pi}{\beta_{E}}t}=\Bigg(\frac{r_c}{r_h}\Bigg)^{-2\Delta}\exp\left(2t\frac{\sqrt{E(1-2E\lambda)}}{1-4\lambda E}\right),\,\quad t \lesssim \beta_{E}\log(S(E))\,.
\end{align}
Hence, the Krylov operator complexity \cite{Rabinovici:2023yex,Xu:2024gfm,Balasubramanian:2024lqk,Heller:2024ldz,Ambrosini:2024sre} exhibits a  dependence on $\lambda$. Now, if we assume that the proposal relating Krylov complexity with that of the ERB  length  $\langle\ell(t)\rangle\sim C_{E}(t)$ \cite{Rabinovici:2023yex,Xu:2024gfm,Balasubramanian:2024lqk,Heller:2024ldz,Ambrosini:2024sre}, is also valid at early times, then it is tempting to conjecture that, $\langle\ell(t)\rangle$ should grow faster in the undeformed theory. However, at late times, we have seen that $\langle\ell(t)\rangle$ saturates faster for certain values of $\lambda$ depending on $\beta$ (see Fig.~(\ref{fig_4})) for the deformed theory. Hence, one can expect that in the intermediate time, there must be a crossover for those values of $\lambda\,.$ However, we note that there is no non-trivial interplay between $\beta$ and $\lambda$ for the early time growth rate of $C_{E}(t)\,.$ Hence, a more careful analysis of $C_{E}(t)$ is required for all time to observe a potential crossover (if any) and subtle interplay between $\beta$ and $\lambda $ in $C_{E}(t)\,.$ This we leave for future investigation, as merely looking at the early time growth of  $C_{E}(t)$, it is not possible to conclude.

\section{Conclusion and Discussion}\label{sec7}
Motivated by the interesting aspects of $T\bar{T}$ deformation, we discuss the emission probability of baby universes (for the good sign of the deformation parameter) after applying such an integrable irrelevant deformation.  We also computed the ERB length growth after $T\bar{T}$ deformation, taking input from the matrix model, and found an interesting effect of inverse temperature on the growth rate (for the bad sign of the deformation parameter).  We also briefly comment on the possible connection with the Krylov complexity.  Also, we have commented on the nature of the matrix model in detail, and also commented on the volume of moduli space (in the Appendix~(\ref{apzb})). Below, we summarize our main findings.
\begin{enumerate}

 \item We find that the emission of baby universes will change in comparison to the undeformed theory if we turn on the Lorentzian evolution.  If we set the duration of the Lorentzian evolution $T\to 0$, both the (undeformed theory and deformed theory) theories have the same emission amplitude for the baby universes.
\item We compute the deformed resolvents and spectral curve and determine the dual matrix model potential for $T\bar{T}$ deformed JT gravity.  While finding the matrix potential for our model, we found that even without adding any instanton corrections, the nature of the potential is such that it leads to less oscillation automatically at some specific value of the energy depending on the value of the deformation parameter and this has a deep connection to eigenvalue repulsion leading to the promisingly different behaviour of the SFF or other quantities like ERB growth, which is worth investigating.  \textcolor{black}{We also comment on the possible connection of Krylov complexity with the growth of the black hole interior.  In that case, to obtain an analytic handle, we only focused on the early time behavior, leaving the detailed study for the future.}

    \item We calculate the growth rate of the ERB for late times.  \textit{Though the classical prediction says that the complexity should increase linearly in time, non-perturbative quantum corrections lead to the saturation of the complexity in JT gravity.  Interestingly, \textcolor{black}{we found that after applying $T\bar T$ deformation, the saturation happens, at least perturbatively in $|\lambda|$.} In conclusion, our analysis reveals a striking dependence of the saturation behavior on the inverse temperature parameter $\beta$.  In particular, we find that for lower values of $\beta$, the JT gravity result exhibits a faster saturation compared to the $T\bar{T}$-deformed theory.  However, this behavior is reversed at higher values,  where the $T\bar{T}$ deformation leads to an earlier saturation.  This qualitative change indicates the presence of a nontrivial crossover in the dynamics}, which can be interpreted as a lower-dimensional analog of a Hawking–Page–type phase transition.  Notably, this feature appears within a perturbative regime associated with the good sign of the deformation parameter $\lambda$, as we have to use input from the matrix model, and we have to work in a regime where we have to make sure that it possesses a one-cut structure.

\item  In Appendix~(\ref{apzb}), we also compute the correction to the moduli space volume arising due to the change in the spectral curve because of the $T\bar{T}$ deformation.  We attempt to find the moduli space volume by taking $\mathcal{O}(\lambda)$ correction to $R_T\,$. We find that due to the presence of the deformation, there is a nontrivial branch cut in the $z-$plane.  We only consider the physical poles contributing to the inverse Laplace integral.  Finally, we show that switching off the deformation would lead to the original volume without the finite cutoff.  We also comment on why the volume of the moduli space should change even if we consider a boundary deformation.  But, one should note that this derivation is valid only for the good sign of $\lambda$. 

\end{enumerate}

\noindent
Now, we end this section by discussing some possible future outlooks. It will be good to do a more systematic study about the crossover as mentioned earlier in future work.
It is known that the $T\bar T$ deformation affects the high-energy levels in a nontrivial way.  Due to this irrelevant deformation, the ultraviolet behaviour of the theory changes.  One interesting extension would be to check the ERB length growth for flat space BMS Schwarzian.  The other quick extension is to calculate the variance of the length in the presence of the $T\bar T$ deformed theory and check at which scale the oscillation starts about the plateau.  One can also try to generalize this computation for self-intersecting geodesics.  As it is still not clearly understood what sort of geometries affect the complexity and lead to the saturation of the interior volume, it is a very interesting direction to analyze even in the absence of $T\bar{T}$ deformation.  \textcolor{black}{Last but not least, it will be interesting to make the connection between the ERB  length  $\langle\ell(t)$ and the Krylov complexity $ C_{E}(t)$ more precise.  So far, this has been established only for  JT gravity by considering contributions from the disk partition function \cite{Rabinovici:2023yex,Ambrosini:2024sre}.  Some progress in this direction has also been made recently for sine-dilaton gravity \cite{Heller:2024ldz}.  If it can be done for our case, then it will serve us with yet another example demonstrating the relationship between ERB length and the Krylov complexity, thereby making the holographic interpretation of Krylov complexity on a stronger footing.} 

\section*{Acknowledgments}
It is a pleasure to thank Edward Witten for commenting on the deformation of the spectral curve and moduli space volume related to our present work. We would also like to thank Nilachal Chakrabarti for discussions on related topics.  AB would like to thank the Department of Physics of BITS Pilani, Goa Campus, for hospitality during the course of this work. S.G (PMRF ID: 1702711) and
S.P (PMRF ID: 1703278) is supported by the Prime Minister’s Research Fellowship of the Government of India. S.G and S.P would like to thank the ``Strings 2025'' organizing committee for giving the opportunity to present posters and NYUAD (New York University Abu Dhabi) for their kind hospitality during the course of the work. AB is supported by the Core Research Grant (CRG/2023/ 001120) by the Department of Science and Technology, Science and Engineering Research Board (India), India. AB also
acknowledges the associateship program of the Indian Academy of Science, Bengaluru.

\appendix

\section{Connected correlation functions of the partition function} \label{aa1}
The $n$-point correlation functions can be written in terms of a sum of topologies with genus $g$ and $n$ boundaries,\vspace{-1cm}\\ 
  \begin{equation}
      \Big\langle \prod_{i=1}^{n} Z(\beta_i)\Big\rangle_{\text{conn}}\equiv \sum_{g=0}^{\infty}\begin{minipage}
          [h]{0.15\linewidth}
	\vspace{1pt}
	\scalebox{2}{\includegraphics[width=\linewidth]{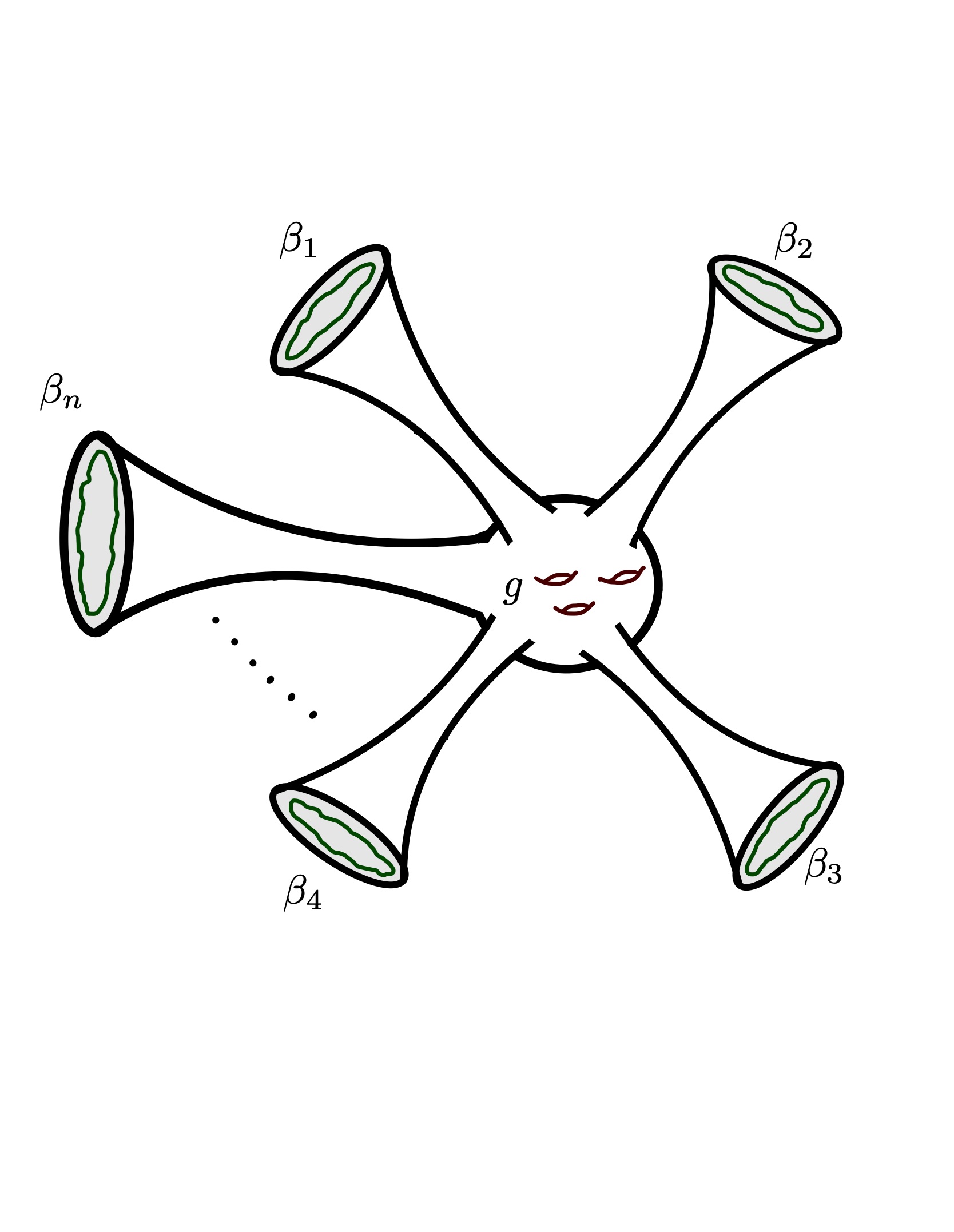}}
      \end{minipage}\hspace{1 cm}\sim \sum_{g=0}^{\infty}\frac{Z_{g,n}(\beta_1,..,\beta_n)}{(e^{S_0})^{2g+n-2}} \,,\label{sum_topology}
      \vspace{-1.5cm}
  \end{equation}
  where $Z_{g,n}(\beta_1,..,\beta_n)$ is given by, 
\begin{equation}
    Z_{g,n}(\beta_1,..,\beta_n)= \int_{0}^{\infty}\prod_{i=1}^{n}b_i db_i\,\mathcal{V}_{g,n}(b_1,..,b_n)Z^{\text{trumpet}}(\beta_i,b_i)\,.
    \label{zgn}
\end{equation}
 Explicitly, the one point correlator $\langle Z(\beta) \rangle$ can be written using \eqref{sum_topology} and \eqref{zgn} as, 
\begin{align}
    \begin{split}
        \big\langle Z(\beta)\big\rangle = e^{S_0} Z^{\text{disk}}(\beta) +\sum_{g=1}^{\infty}e^{(1-2g)S_0}\int_{0}^{\infty}db\,b\,\mathcal{V}_{g,1}(b) Z^{\text{trumpet}}(\beta,b)\,.
    \end{split}
\end{align}
Considering contributions up to $g=1$,
\begin{equation}
    \big\langle Z(\beta)\big\rangle = e^{S_0} Z^{\text{disk}}(\beta) +e^{-S_0}
    \int_{0}^{\infty}db\,b\,\mathcal{V}_{1,1}(b) Z^{\text{trumpet}}(\beta,b) + \mathcal{O}(e^{-3S_0})    
\end{equation}
where $\mathcal{V}_{1,1}(b)$ is given as
\begin{equation}
    \mathcal{V}_{1,1}(b) = \frac{1}{48} (b^2 + 4\pi^2)\,.
\end{equation}
For pure JT gravity, we have,
\begin{equation}
    \big\langle Z(\beta)\big\rangle = e^{S_0} Z^{\text{disk}}(\beta) +e^{-S_0} \frac{\sqrt{\beta}}{12\sqrt{\pi}}(\pi^2  + \beta)  + \mathcal{O}(e^{-3S_0})\,.
\end{equation}
Similarly, the connected two-point correlator $\langle Z(\beta_1)Z(\beta_2)\rangle_{\text{conn}}$ in pure JT gravity can be obtained as follows,\\
\begin{align}
    \begin{split}
        \big \langle Z(\beta_1)Z(\beta_2)\big\rangle_{\text{conn}} = Z_{0,2}(\beta_1,\beta_2) + \sum_{g=1}^\infty e^{-2gS_0} \int_0^\infty\prod_{i=1}^2 b_idb_i \mathcal{V}_{g,2}(b_1,b_2) Z^{\text{trumpet}}(\beta_i,b_i)\,.
    \end{split}
\end{align}
Again, considering the contribution up to $g=1$, we have,
\begin{align}
    \begin{split}
        \big \langle Z(\beta_1)Z(\beta_2)\big\rangle_{\text{conn}} = \frac{\sqrt{\beta_1\beta_2}}{2\pi(\beta_1 + \beta_2)} + e^{-2S_0} \frac{\sqrt{\beta_1\beta_2}}{12\pi}\bigg[3\pi^4 + 4\pi^2(\beta_1+\beta_2) + 2(\beta_1^2 + \beta_1\beta_2+\beta_2^2)\bigg] +\mathcal{O}(e^{-4S_0})
    \end{split}
\end{align}
where we used 
\begin{equation}
    \mathcal{V}_{1,2}(b_1,b_2) = \frac{1}{192}(4\pi^2 + b_1^2 + b_2^2)(12\pi^2 + b_1^2 + b_2^2)\,.
\end{equation}
One can extend these for the $T\bar{T}$ deformed case. \\\\
\textbf{$T\bar{T}$ deformed correlation functions :} We can perform similar computations using the deformed partition functions to obtain the deformed one-point correlator,
\begin{equation}
    \big\langle Z(\beta,\lambda)\big\rangle = e^{S_0} Z^{\text{disk}}_\lambda(\beta) +e^{-S_0}
    \int_{0}^{\infty}db\,b\,\mathcal{V}_{1,1}(b) Z^{\text{trumpet}}_\lambda(\beta,b) + \mathcal{O}(e^{-3S_0})\,.  
\end{equation}
Using the form of $Z^{\text{trumpet}}_\lambda(\beta,b)$ in equation \eqref{5.9u} we have
\begin{align}
    \begin{split}
        \big\langle Z(\beta,\lambda)\big\rangle = e^{S_0} Z^{\text{disk}}_\lambda(\beta) +e^{-S_0} \frac{\sqrt{\beta}}{12\sqrt{\pi}}(\pi^2  + \beta)  +e^{-S_0}\lambda \frac{(\pi^2 - 3\beta)}{24\sqrt{\pi\beta}}  + \mathcal{O}(e^{-3S_0})\label{3.52k}\,. 
    \end{split}
\end{align}
The one-point function of the partition function in \eqref{3.52k} is perturbative in $\lambda$. Furthermore, the $T\bar{T}$ deformed moduli space volumes (for the good sign of the deformation parameter) have been calculated in Appendix~(\ref{apzb}). As the spectral curve changes non-trivially after applying $T\bar{T}$ deformation, we also expect the moduli space volume to change. It is an aspect of further investigation to know why the moduli space volume, being a bulk quantity, will change despite applying the integrable deformation to the boundary theory. We also comment on this question in Appendix~\eqref{apzb}.

\section{\bf $T\bar{T}$ deformation of the volume of moduli space}\label{a1}
\label{apzb}
We begin by introducing the spectral curve.  We define the following curve,
$$E(z)=-z^2\,.$$
In order to find the  resolvents for the trumpet partition function, we do the following integral transform \cite{Stanford:2019vob},
\begin{align}
\begin{split}
    R_{T}&=-\int_0^\infty d\beta e^{-\beta z^2}Z_{trumpet}^{\lambda} (\beta,b)\,,\\&
    =\frac{1}{2}\, \lambda\,  z \,e^{-b z} (b z-3)-\frac{e^{-b z}}{2 z},\,\quad \lambda<0\,.
    \end{split}
\end{align}

\noindent
Now, the resolvents can be written in terms of moduli space volumes as follows \cite{Stanford:2019vob},
\begin{align}
R_{g}(-z_1^2,\cdots ,-z_n^2)=\int_0^\infty V_g(b_1,\cdots b_n)\prod_{j=1}^n b_j db_j \,R_T(b_j,z_j;\lambda)\,.
\end{align}
Hence, we can write the deformed volume as
\begin{align}
  V_g(b_1,\cdots b_n)_{\lambda} =\int_{c_0+i \mathbb{R}}R_g^\lambda(-z_1^2,\cdots ,-z_n^2) \prod_{j=1}^n \frac{dz_j}{2\pi i }\,\,\frac{1}{R_T(b_j,z_j;\lambda) b_j}
\end{align}
For our case, we specifically compute the following, 
\begin{align}
    V_1(b_1)_{\lambda}=\int_{c_0+i \mathbb{R}} \,\, \frac{dz}{2\pi i }\,\,\frac{1}{R_T(b,z;\lambda) b} \,\,R_1^\lambda(-z^2)\,.
\end{align}
\noindent
Now we can define \begin{align}
\begin{split}
   & W_{g,n}(z-1,z_2,\cdots,z_n;\lambda)=\textrm{Res}_{z\rightarrow 0}\mathcal{K}(z_1,z;\lambda)\Bigg[W_{g-1,n+1}(z,-z,z_2\cdots,z_n;\lambda)+\\&\hspace{9cm}\sum_{\substack{h_1+h_2=g\\I_1\bigcup I_2=J}}^*W_{h_1,1+|I_1|}(z,I_1,\lambda)\,\,W_{h_2,1+|I_2|}(-z,I_2,\lambda)\Bigg]\,
    \end{split}
\end{align}
\noindent
Here the `$\textbf{*}$' indicates that one should not consider cases when $\{h_2,I_2\} \mbox{ or } \{h_1,I_1\}=\{g,J\}$.
\noindent
The kernel is given by the following expression,
\begin{figure}[t!]
\centering
\scalebox{0.27}{\includegraphics{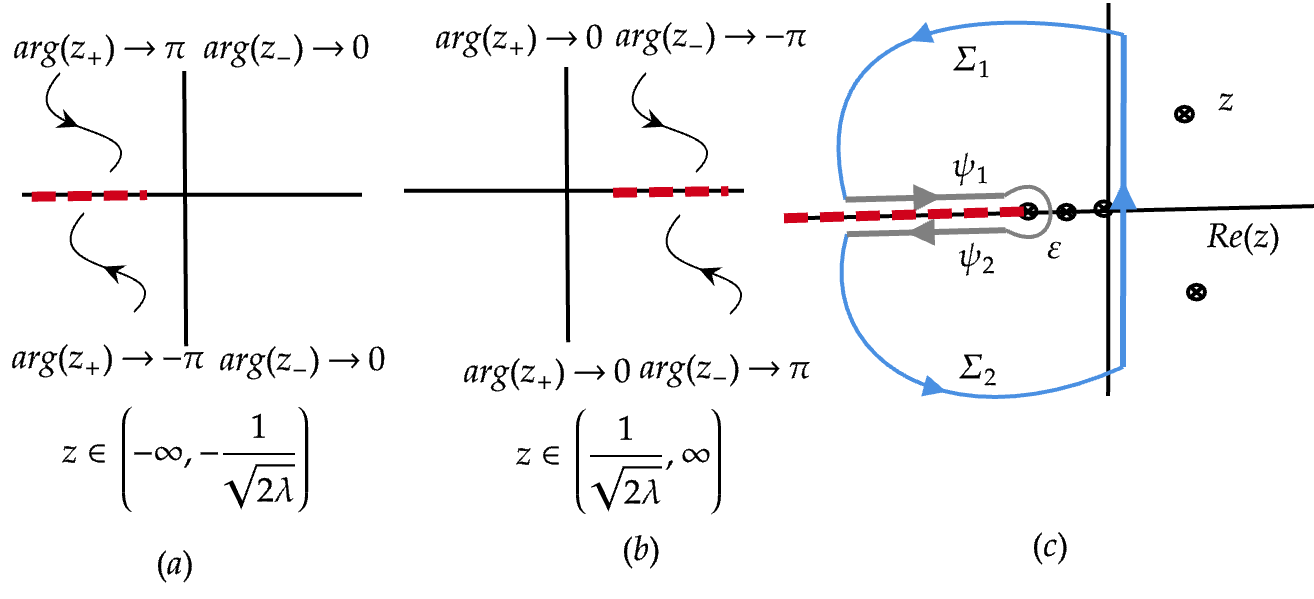}}
\caption{Contour for the integral to find the moduli space volume in $T\bar{T}$ deformed case ($z_+=z+\frac{1}{\sqrt{2\lambda}},\,\,z_{-}=\frac{1}{\sqrt{2\lambda}}-z$). }\label{fig5}
\end{figure}
\begin{align}
\begin{split}
    \mathcal{K}(z_1,z;\lambda)&=\frac{1}{2[W_{0,1}(z;\lambda)+W_{0,1}(-z,\lambda)]}\int_{-z}^{z}
dz_2 W_{0,2}(z_1,z_2,\lambda)\\&
=\frac{(2+\lambda z_1^2)\sqrt{4+\lambda z^2}}{(2+\lambda z^2)\sqrt{4+\lambda z_1^2}}\frac{4\pi \textrm{ cosec}(\pi z\sqrt{4+\lambda z^2})}{(z_1^2-z^2)[4+\lambda(z_1^2+z^2)]}\,.
\end{split}
\end{align}
\noindent
Now, by change of variables $R_{g,n}$'s  are related with $W_{g,n}$'s in the following way \cite{Mirzakhani:2011gta,Mirzakhani:2006eta},
\begin{align}
   W_{g,n}(z_1,z_2,\cdots z_n;\lambda) = R_{g,n}(E(z_1),\cdots, E(z_n);\lambda) \,\,E'(z_1)\cdots E'(z_n)\,.
\end{align}
Now following \cite{Griguolo:2021wgy}, \begin{align}
    W_{1,1}(z_1,\lambda)=\frac{(2+8\lambda z_1^2)[6+\pi^2 z_1^2(4+8\lambda z_1^2)]}{3z_1^4(4+8\lambda z_1^2)^{5/2}}\,.
\end{align}

Hence,
\begin{align}
   R_1^\lambda(-z_1^2) = \,R_{1,1}(E(z_1);\lambda)=-\frac{(2+8\lambda z_1^2)[6+\pi^2 z_1^2(4+8\lambda z_1^2)]}{6z_1^5(4+8\lambda z_1^2)^{5/2}}\,,
\end{align}\\

Therefore,\\
\hfsetfillcolor{gray!8}
\hfsetbordercolor{white}
\begin{align}
\begin{split}
\tikzmarkin[disable rounded corners=true]{rey1}
(4.2,-0.7)(-0.3,1.0) 
V_1(b)_{\lambda}&=\int_{C_0+i\mathbb{R}} \,\, \frac{dz_1}{2\pi i }\,\,\frac{1}{R_T(b,z_1;\lambda) b} \,\,\Bigg[-\frac{(2+8\lambda z_1^2)[6+\pi^2 z_1^2(4+8\lambda z_1^2)]}{6z_1^5(4+8\lambda z_1^2)^{5/2}}\Bigg]\\&
    =-\frac{1}{b}\oint_{\mathbb{C}} \frac{dz_1}{2\pi i}\frac{e^{b z_1} \left(\lambda  z_1^2+2\right) \left(\pi ^2 z_1^2 \left(8\lambda  z_1^2+4\right)+6\right)}{3 z_1^4 \left(8\lambda  z_1^2+4\right){}^{5/2} \left(b \lambda  z_1^3-3 \lambda  z_1^2-1\right)}+\cdots\,.
\tikzmarkend{rey1}
\end{split}
\end{align}\\
\noindent
Now for good sign of deformation parameter i.e. $\lambda<0\to -|\lambda|$ \footnote{In the subsequent computation we use $|\lambda|\to \lambda\,.$}, the following integral to be done :
\begin{align}
    \begin{split}
        \Theta_{1}(b)=\int_{-i\infty}^{i\infty}\frac{dz}{2\pi i}\frac{e^{b z} \left(2-8 \lambda  z^2\right) \left(\pi ^2 z^2 \left(4-8 \lambda  z^2\right)+6\right)}{3 z^4 \left(4-8 \lambda  z^2\right)^{5/2} \left(-b \lambda  z^3+3 \lambda  z^2-1\right)}=\int_{-i\infty}^{i\infty} \frac{dz}{2\pi i}f(z) dz\label{B.11}
    \end{split}
\end{align}
where,
\begin{align}
\begin{split}
    &f(z)=g(z)\,\left(\frac{1}{\sqrt{2 \lambda }}+z\right)^{-5/2} \left(\frac{1}{\sqrt{2 \lambda }}-z\right)^{-5/2},\,\\&\textrm{with},\,  g(z)=(8\lambda)^{-5/2}\frac{e^{b z} \left(2-8 \lambda  z^2\right) \left(\pi ^2 z^2 \left(4-8 \lambda  z^2\right)+6\right)}{3 z^4 (-b \lambda z^3+3\lambda z^2-1)}\,.
    \end{split}
\end{align}
The integral in \eqref{B.11} can be decomposed as,
\begin{align}
    \begin{split}
       \int_{-i\infty}^{i\infty} \frac{dz}{2\pi i}f(z)&=\oint_{C} \frac{dz}{2\pi i}f(z)-\cancel{\int_{\Sigma_1+\Sigma_2}\frac{dz}{2\pi i}f(z)}-\int_{\psi_1+\psi_2}\frac{dz}{2\pi i}f(z)-\int_{\epsilon}\frac{dz}{2\pi i}f(z)\,,\\ &
       =\sum \textrm{residues @ poles}- \frac{1}{2\pi i}\textrm{Disc.}[f(z)|\psi_1,\psi_2]-\int_{\epsilon}\frac{dz}{2\pi i}f(z)\,.
    \end{split}
\end{align}
Analyzing the singularity structure of the function $f(z)$ one can see that the function has poles at, $z=0,\,z_{r},z_c,\bar{z}_c$, where,
\begin{align}
    \begin{split}
     &   z_r=-\frac{\sqrt[3]{b^2 \lambda ^2+\sqrt{b^4 \lambda ^4-4 b^2 \lambda ^5}-2 \lambda ^3}}{\sqrt[3]{2} b \lambda }-\frac{\sqrt[3]{2} \lambda }{b \sqrt[3]{b^2 \lambda ^2+\sqrt{b^4 \lambda ^4-4 b^2 \lambda ^5}-2 \lambda ^3}}+\frac{1}{b}\,,\\ &
     z_c=\frac{\left(1-i \sqrt{3}\right) \sqrt[3]{b^2 \lambda ^2+\sqrt{b^4 \lambda ^4-4 b^2 \lambda ^5}-2 \lambda ^3}}{2 \sqrt[3]{2} b \lambda }+\frac{\left(1+i \sqrt{3}\right) \lambda }{2^{2/3} b \sqrt[3]{b^2 \lambda ^2+\sqrt{b^4 \lambda ^4-4 b^2 \lambda ^5}-2 \lambda ^3}}+\frac{1}{b}\,.
    \end{split}
\end{align}
One can easily see that the complex poles do not lie inside the chosen contour \footnote{These complex poles are unphysical since the residue from these two poles will give a divergent contribution if we set the deformation parameter $\lambda$ to be zero. So, we only take the contribution from the physical poles such that we have a smooth $\lambda\to0$ limit.}, as shown in Fig.~(\ref{fig5}). Then, only the residue from $z=0,\,z_r$ contributes to the sum of residues. Hence,
\begin{align}
    \sum \textrm{residues @ poles}=-\frac{1}{48} b \left(b^2+18 \lambda +4 \pi ^2\right)+e^{-\frac{\sqrt[3]{\frac{\zeta }{\lambda }}}{\sqrt[3]{2}}-\frac{\sqrt[3]{2} \sqrt[3]{\lambda }}{\sqrt[3]{\zeta }}}\frac{N}{D}
\end{align}
with,
\begin{align}
\begin{split}
    N=&-8  \big(b^2 \left(-2 \sqrt[3]{2} \sqrt[3]{\zeta } \lambda ^{5/3}+\left(\zeta  \lambda ^2\right)^{2/3}+4\ 2^{2/3} \lambda ^2\right)-12 \zeta ^{2/3} \lambda ^{7/3}+2 \sqrt[3]{2} \sqrt[3]{\zeta } \lambda ^{5/3} (6 \lambda -\chi )\\ &+2^{8/3} \lambda ^2 (\chi -3 \lambda )\big) \big(b^6 \lambda ^3 \left(3 \sqrt[3]{\zeta } \lambda ^{2/3}+2 \sqrt[3]{2} \pi ^2\right)\\&+b^4 \big(2 \pi ^2 \lambda ^3 \left(-2\ 2^{2/3} \zeta ^{2/3} \sqrt[3]{\lambda }+11 \sqrt[3]{\zeta } \lambda ^{2/3}+\sqrt[3]{2} (\chi -26 \lambda )\right)\\ &\hspace{-0.7 cm}+3 \sqrt[3]{\zeta } \lambda ^{11/3} (\chi -2 \lambda )\big)+2 \pi ^2 b^2 \lambda ^{10/3} \left(2^{2/3} \zeta ^{2/3} (23 \lambda -2 \chi )-76 \sqrt[3]{\zeta } \lambda ^{4/3}+11 \chi  \sqrt[3]{\zeta  \lambda }-24 \sqrt[3]{2} \lambda ^{2/3} \chi +118 \sqrt[3]{2} \lambda ^{5/3}\right)\\ &+36 \pi ^2 \lambda ^{13/3} \left(2^{2/3} \zeta ^{2/3} (\chi -3 \lambda )+6 \sqrt[3]{\zeta } \lambda ^{4/3}-3 \chi  \sqrt[3]{\zeta  \lambda }+4 \sqrt[3]{2} \lambda ^{2/3} \chi -6 \sqrt[3]{2} \lambda ^{5/3}\right)\big)
    \end{split}
\end{align}
and,
\begin{align}
    \begin{split}
        D=&\frac{9}{b\zeta}\left(-2 \sqrt[3]{\zeta } \lambda^{5/3}+2^{2/3} \left(\zeta  \lambda ^2\right)^{2/3}+2 \sqrt[3]{2} \lambda^2\right)^5 \left(2 \sqrt[3]{\zeta } \lambda ^{5/3}+2^{2/3} \left(\zeta  \lambda ^2\right)^{2/3}+2 \sqrt[3]{2} \lambda ^2\right)\\ & \times \Bigg[b^4 \lambda +b^2 \left(\lambda  \left(2\ 2^{2/3} \sqrt[3]{\zeta  \lambda ^2}+\chi \right)-\sqrt[3]{2} \left(\zeta  \lambda ^2\right)^{2/3}-8 \lambda ^2\right)+6 \sqrt[3]{2} \zeta ^{2/3} \lambda ^{7/3}+\\& 2\ 2^{2/3} \sqrt[3]{\zeta } \lambda^{5/3} (\chi -3 \lambda)-\sqrt[3]{2} \chi  (\zeta  \lambda ^2)^{2/3}+12 \lambda ^3-6 \lambda ^2 \chi\Bigg]^{5/2}
    \end{split}
\end{align}
where, $\chi=b^4-4 b^2 \lambda,\,\zeta=b^2-2 \lambda+\chi$.
\\\\
Further, the discontinuity along the branch cut is,
\begin{align}
    \begin{split}
        \textrm{Disc.}[f(z),\psi_1,\psi_2]=2i \int ^{-\infty}_{-\frac{1}{\sqrt{2\lambda}}}dt\,g(t) \left|\frac{1}{\sqrt{2 \lambda }}+t\right|^{-5/2} \left|\frac{1}{\sqrt{2 \lambda }}-t\right|^{-5/2}
    \end{split}
\end{align}
and the integral over the small circle becomes,
\begin{align}
    \begin{split}
\int _{\varepsilon}&f(z) dz\to \frac{i \lambda ^{3/4} e^{-\frac{b}{\sqrt{2} \sqrt{\lambda }}} \left(2 \sqrt{2} \left(3 b^2+4 \pi ^2\right) \sqrt{\lambda }+63 b \lambda +8 \pi ^2 b+87 \sqrt{2} \lambda ^{3/2}\right)}{6\times\ 2^{3/4} b \,\sqrt{\varepsilon } \left(\sqrt{2} b+2 \sqrt{\lambda }\right)^2}\\&\hspace{6 cm}-\frac{i \lambda ^{5/4} e^{-\frac{b}{\sqrt{2} \sqrt{\lambda }}}}{3\times\ 2^{3/4} b\, \varepsilon ^{3/2} \left(\sqrt{2} b+2 \sqrt{\lambda }\right)}\,.
    \end{split}
\end{align}
One can notice that this integral is purely divergent after taking the limit $\varepsilon\to 0$. However, for undeformed case ($\lambda=0$), the exponential term falls off faster and makes the integral zero, which is expected, and the divergence comes from purely the branch point of the integrand and is non-perturbative (in the integrated, only the contribution from $R_T$ is perturbative in $\lambda$ and we take up to $\mathcal{O}(\lambda)$.). Then, we propose the regularized volume as,

\begin{align}
    \begin{split}
    \boxed{ \textrm{Reg.}[ V_{1}(b)]=V_1{(b)}-\frac{1}{b}\int_{\varepsilon} \frac{dz}{2\pi i}f(z)\,.}\label{B.23}
    \end{split}
\end{align}
We would like to emphasize that the definition of regularized volume in \eqref{B.23} is purely non-perturbative in $\lambda $ because if one adds higher order correction of $\lambda $ in $R_T$, the pole structure only changes (there will be more poles) but the branch cut structure will be unchanged.
Now, with definition in \eqref{B.23} the volume becomes,\\
\hfsetfillcolor{gray!12}
\hfsetbordercolor{white}
\begin{align}
    \begin{split}
    \tikzmarkin[disable rounded corners=true]{rey2}
(1,-1.2)(-0.3,1.0) 
       \textrm{Reg.}[V_{1}(b)]=\frac{1}{48}  \left(b^2+4\pi^2+18 \lambda \right)-\frac{{1}}{b}e^{-\frac{\sqrt[3]{\frac{\zeta }{\lambda }}}{\sqrt[3]{2}}-\frac{\sqrt[3]{2} \sqrt[3]{\lambda }}{\sqrt[3]{\zeta }}}\frac{N}{D}-\frac{1}{\pi\,b}\int_{-\frac{1}{\sqrt{2\lambda}}}^{-\infty}\left|t^2-\frac{1}{2\lambda}\right|^{-5/2}\,g(t),\,\,\lambda>0\,.
        \tikzmarkend{rey2}        \end{split}
        \label{B.24}
\end{align}\\
One can easily verify that by taking $\lambda\to0$ limit, \eqref{B.24} reproduces the undeformed volume which is $V_{1}(b)=\frac{1}{48}(b^2+4\pi^2)$. 
\begin{align}
    f(z)=g(z) \left|\frac{1}{\sqrt{2 \lambda }}+z\right|^{-5/2}\left|\frac{1}{\sqrt{2 \lambda }}-z\right|^{-5/2} e^{-i\,5/2\,\textrm{arg}(\frac{1}{\sqrt{2 \lambda }}+z)-i\,5/2\,\textrm{arg}(\frac{1}{\sqrt{2 \lambda }}-z)}\end{align}
Now, an immediate question arises as to how a typical boundary deformation could lead to a change in the volume of the moduli space, which is a bulk quantity. In the next subsection, we intend to give a flavor of how the boundary deformation has a non-trivial back reaction in bulk.
\section*{A hint towards the deformed volume} \label{b1}
We computed above the deformed moduli space volume for $T\bar{T}$ deformed Schwarzian theory as the spectral curve changes. We found that $W_{0,1}, W_{0,2}$ do not change in comparison to JT gravity. But $W_{1,1}$ changes even if we choose variables wisely.
\noindent
For a general spectral curve $(S_a,\mathbb{C},x,y)$, we review in this section briefly how to obtain $W_{g,n}$ from the spectral curve itself.
\noindent
The formula  to obtain the \textbf{symplectic invariant descendants}  is given by \cite{Eynard:2011kk},
\begin{align}
    W_{g,n}(S_a;z_1,z_2\cdots, z_n)=2^{d_{g,n}}\sum_{d_1+d_2+\cdots+d_n\leq d_{g,n}}\prod_id\xi_{d_i}(z_i)\Bigg\langle e^{\frac{1}{2}\sum_{\delta}l_{\delta_{*}}\hat{B}(\psi,\psi')}e^{\sum_{k}\tilde{t}_k\kappa_k}\prod_i \psi_i^{d_i}\Bigg\rangle_{g,n}
\end{align}

where, \begin{align}
    \Big\langle\tau_{d_1}\tau_{d_2}\cdots \tau_{d_n}\Big\rangle_{g}=\int_{\bar{\mathcal{M}}_{g,n}}\psi_1^{d_1}\psi_2^{d_2}\psi_3^{d_3}\cdots \psi_{n}^{d_n}\hspace {2 cm} \textrm{with} \,\,\,d_i\geq 0
\end{align} along with,  $d_{g,n}=3g-3+n$.
For this correlator to be non vanishing, $$d_1+d_2+d_3+\cdots +d_n=3g-3+n=\textrm{dim}(\mathcal{M}_{g,n})\,.$$
In the above notation for tautological $\psi_i$ classes, which are defined as the {\textit {first Chern classes}} of the canonical section of the relative dualizing sheaf \footnote{$\omega_{\pi}$ relative dualizing sheaf with $\psi_i=c_1(\sigma_i^*(\omega_\pi))$.} corresponding to the {forgetful map}, $$\pi:\bar{\mathcal{M}}_{g,n+1}\rightarrow \bar{\mathcal{M}}_{g,n}$$  where, $\psi^{d_i}=\tau_{d_i}$. The
\textit{`times'} are defined using the Laplace transform of the one-form  $ydx$ along some steepest descent curve from the branch point \footnote{ The branch point is the zero of $dx(z)$. For our case, we found $a=0$.} $a$ to $\infty$.

\begin{align} \label{A.25}
    e^{-\sum_{k} \tilde{t}_ku^{-k}}=\frac{2u^{3/2}e^{ux(a)}}{\sqrt{\pi}}\int_{\gamma}e^{-ux }y dx\,.
\end{align}
\textbullet \,The \textit{one-forms} are defined as,
\begin{align}
d\xi_{d}=-\textrm{Res}_{z'\rightarrow a}\mathcal{B}(z,z')\frac{(2d-1)!!}{2^d(x(z')-x(a))^{d+1/2}}
\end{align}
where,
\begin{align}
    \mathcal{B}(z_1,z_2)\xrightarrow[z_1\rightarrow z_2]{}\frac{dz_1\otimes dz_2}{(z_1-z_2)^2}+\cdots(\textrm{Holomorphic non-singular terms})
\end{align}
is a symmetric 2nd kind of differential with no other pole except the double pole.

\noindent
\textbullet \,\,The other quantity $
\hat{\mathcal{B}}$ is defined as,
\begin{align}
\hat{\mathcal{B}}=\sum_{k,l}\hat{\mathcal{B}}_{k,l}\psi^k\psi'^l 
\end{align}\\ is defined by the double Laplace transform of the {\textit{ Bergman kernel}}.
\begin{align}
\sum_{k,l}\hat{\mathcal{B}}_{k,l}\psi^k\psi'^l=\frac{(uu')^{1/2}}{2\pi}e^{(u+u')x(a)}\int_{z\in\gamma}\int_{z'\in\gamma}e^{-ux(z)-u'x(z')}\Big(\mathcal{B}(z,z')-{{\mathcal{C}}(z_1,z_2)}\Big)
\end{align}
\noindent
where $\mathcal{C}(z_1,z_2)$is the trivial part of the double pole, and it is defined as,

\begin{align}
    \mathcal{C}(z_1,z_2)=\frac{dx(z_1)\otimes dx(z_2)}{4\sqrt{x(z_1)-x(a)}\sqrt{x(z_2)-x(a)}}\,\,\frac{1}{\Big(\sqrt{x(z_1)-x(a)}-\sqrt{x(z_2)-x(a)}\Big)^2}\,.
\end{align}
The $\sum_{\delta}$ means that we have to take care of the sum over all the \textit{boundary divisors} and $l_{\delta^*}$ is the operator pinching of the specific boundary circle $\delta$ maintaining the stability of the graphs. $\psi,\psi'$ are the first Chern classes of the cotangent line bundle corresponding to the nodal point.
Now, for our case, the spectral curve is given by, 
\begin{align}
    y(x)=\frac{(1-4\lambda x)}{4\pi^2}\textrm{sinh}\Big(2\pi \sqrt{x(1-2\lambda x)}\Big)
\end{align}
and $W_{1,1}$ (which is one of the main ingredients of computing the volume) needs two\textit{ one-forms} defined above i.e $d\xi_0$ and $d\xi_1$. \textit{We found that $d\xi_1$  is dependent on $\lambda$, but $d\xi_0$ is not, implying that the change in the volume of the moduli space volume takes place for $V_{1,1}(b)$, which agrees with our above computation.} The differentials are given by,

\begin{align}
    d\xi_0\sim \frac{1}{z}\hspace{1.6 cm}\textrm{ and  }\hspace{2 cm}d\xi_1\sim \frac{1}{z^3}-\frac{3 \lambda }{z}.
\end{align}
The one form $ydx$ remains unchanged by choosing $x(1-2\lambda x)=z^2$ and the `times' ($\tilde{t}_k$) also remain unchanged as the integral path in \eqref{A.25} remains the same even for the deformed case.

\bibliography{ref1}
\bibliographystyle{jhep}
\end{document}